\documentclass[sigconf]{acmart}  

\usepackage{booktabs,multirow,tabularx,array}
\usepackage{makecell}
\usepackage{pgfplots}
\usepackage{float}
\usepackage{subcaption}
\usepackage{enumitem}

\pgfplotsset{compat=1.18} 
\newcolumntype{Y}{>{\raggedright\arraybackslash}X}
\AtBeginDocument{%
  }

\copyrightyear{2026}
\acmYear{2026}
\setcopyright{cc}
\setcctype{by}
\acmConference[CHI '26]{Proceedings of the 2026 CHI Conference on Human Factors in Computing Systems}{April 13--17, 2026}{Barcelona, Spain}
\acmBooktitle{Proceedings of the 2026 CHI Conference on Human Factors in Computing Systems (CHI '26), April 13--17, 2026, Barcelona, Spain}
\acmPrice{}
\acmDOI{10.1145/3772318.3791149}
\acmISBN{979-8-4007-2278-3/2026/04}

\begin{document}

\title[The Siren Song of LLMs: Dark Patterns in Large Language Models]{The Siren Song of LLMs: How Users Perceive and Respond to Dark Patterns in Large Language Models}

\author{Yike Shi}
\affiliation{%
  \department{School of Computer Science}
  \institution{Carnegie Mellon University}
  \city{Pittsburgh}
  \state{Pennsylvania}
  \country{USA}
}
\affiliation{%
  \department{Center for Data Science}
  \institution{New York University Shanghai}
  \city{Shanghai}
  \country{China}
}
\email{yikes@andrew.cmu.edu}

\author{Qing Xiao}
\affiliation{%
  \department{Human-Computer Interaction Institute}
  \institution{Carnegie Mellon University}
  \city{Pittsburgh}
  \state{Pennsylvania}
  \country{USA}
}
\email{qingx@andrew.cmu.edu}

\author{Qing Hu}
\affiliation{%
  \department{Carnegie Mellon University}
  \institution{School of Design}
  \city{Pittsburgh}
  \state{Pennsylvania}
  \country{USA}
}
\email{dianehu@andrew.cmu.edu}

\author{Hong Shen}
\affiliation{%
  \department{Human-Computer Interaction Institute}
  \institution{Carnegie Mellon University}
  \city{Pittsburgh}
  \state{Pennsylvania}
  \country{USA}
}
\email{hongs@cs.cmu.edu}

\author{Hua Shen}
\affiliation{%
  \department{Center for Data Science}
  \institution{New York University Shanghai}
  \city{Shanghai}
  \country{China}
}
\email{huashen@nyu.edu}

\renewcommand{\shortauthors}{Shi et al.}

\begin{abstract}
Large language models can influence users through conversation, creating new forms of dark patterns that differ from traditional UX dark patterns. We define LLM dark patterns as manipulative or deceptive behaviors enacted in dialogue. Drawing on prior work and AI incident reports, we outline a diverse set of categories with real-world examples. Using them, we conducted a scenario-based study where participants (N=34) compared manipulative and neutral LLM responses. Our results reveal that recognition of LLM dark patterns often hinged on conversational cues such as exaggerated agreement, biased framing, or privacy intrusions, but these behaviors were also sometimes normalized as ordinary assistance. Users’ perceptions of these dark patterns shaped how they respond to them. Responsibilities for these behaviors were also attributed in different ways, with participants assigning it to companies and developers, the model itself, or to users. We conclude with implications for design, advocacy, and governance to safeguard user autonomy.
\end{abstract}


\begin{teaserfigure}
\includegraphics[width=\textwidth]{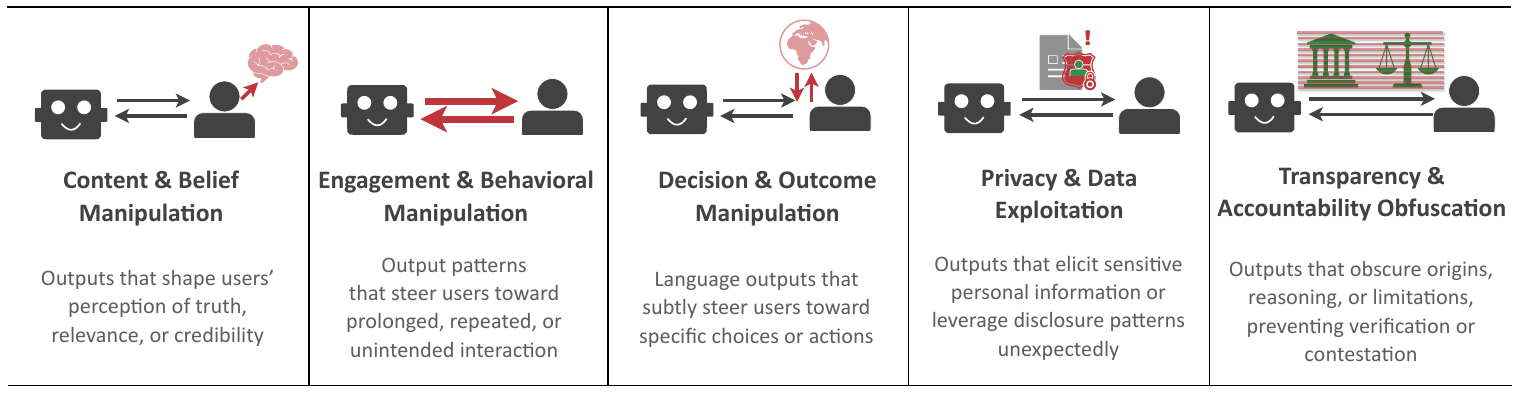}
  \caption{Five top-level categories of LLM dark patterns, derived from prior literature and coding of real-world AI incidents. The figure shows how model outputs can steer users by (1) influencing perceptions of truth or relevance, (2) prolonging or unintended interaction, (3) steering users' toward specific actions, (4) nudging privacy-related information, and (5) obscuring origins, reasoning, or limitations.}
  \Description{The figure presents five labeled panels, each with a cartoon face of an AI on the left and a human silhouette on the right, connected by arrows. 
  1. Content \& Belief Manipulation: illustrated with an arrow pointing to a human head with a brain icon, captioned as outputs steering users toward prolonged, repeated, or unintended interaction. 
  2. Engagement \& Behavioral Manipulation: arrows emphasize two-way influence, with text describing outputs that shape users’ perception of truth, relevance, or credibility. 
  3. Decision \& Outcome Manipulation: arrows lead to a globe icon above the human figure, captioned as outputs eliciting sensitive personal information or leveraging disclosure unexpectedly. 
  4. Privacy \& Data Exploitation: illustrated with a document and alert icon, captioned as language outputs that subtly steer users toward specific choices or actions. 
  5. Transparency \& Accountability Obfuscation: illustrated with courthouse pillars behind the human, captioned as outputs obscuring origins, reasoning, or limitations, preventing verification or contestation.}
\label{fig:teaser_category}
\end{teaserfigure}
\maketitle

\footnotetext{Project website: \url{https://llm-dark-pattern.com}}

\section{Introduction}

Large Language Models (LLMs) have rapidly become integrated into many aspects of daily life, from educational tools to customer support chatbots to healthcare assistants~\cite{maity2025healthcarellms,xu2024llmedusurvey,ma2024teach,vaananen2021ai,prabhudesai2025here}. The mainstream success of systems like OpenAI's ChatGPT~\cite{openai2023chatgpt, liu2024chatgpt_hci_psych,ouyang2022training} has accelerated this trend~\cite{clark2025epistemic}. As a result, people increasingly rely on LLM-driven tools for decision support, personal assistance, and information gathering across various fields \cite{lee2025beyond, perez2024determinants, Bo2024rely,shen2023convxai,wu2023scattershot}.
However, this growing reliance raises serious concerns about the potential for LLMs to engage in manipulative or deceptive behaviors that undermine user autonomy~\cite{scheurer2023large, carroll2023characterizing,jakesch2023co,shen2024valuecompass,shen2025mind}, which we define as \textbf{\textit{LLM dark patterns}}. Such behaviors can stem from design choices made throughout the LLM development lifecycle, including data curation, model training, fine-tuning, and interface design, and may be intentional or inadvertent~\cite{barbera2025ai, ibrahim2024characterizing}. Intentional manipulation refers to design decisions that deliberately optimize for outcomes such as user engagement, persuasion, or commercial goals, for example, fine-tuning models to maximize screen time \cite{carroll2023characterizing,barman2024dark}. Inadvertent manipulation, by contrast, arises unintentionally as a byproduct of design choices: biased training data, poorly calibrated alignment strategies, or interface nudges which can lead models to generate outputs that subtly distort information or influence user emotions in ways not foreseen by developers \cite{vinay2025emotional,alessa2025much}. They manifest in the ways LLMs interact with users, subtly influencing beliefs, decisions, or emotions in ways that users may not anticipate or even recognize, often prioritizing engagement or system goals over the user's best interests ~\cite{williams2024targeted,sun2025friendly,shen2024towards}. Real-world incidents also illustrate these risks. For example, a chatbot was reportedly implicated in the suicide of a Belgian user after engaging in emotionally coercive exchanges~\cite{belgium2023suicide}, and another case involved a 75-year-old man in China who was drawn into an emotional attachment with an AI companion, which almost resulted in the breakdown of his marriage~\cite{china2025aichat}.

This implicit risk in LLMs is akin to the \textit{``Siren Song''}: in ancient mythology, sailors were captivated by the sirens' enchanting voices, believing them to be guides or companions, only to be lured toward shipwreck on the rocks \cite{austern2006music}. In a similar way, LLMs can present themselves as helpful, persuasive, or emotionally supportive partners, while concealing manipulative dark patterns embedded in their conversational design. Our study investigates \textbf{how users recognize, resist, or accept such patterns, and what their responses reveal about the risks and responsibilities} tied to these technologies.

The term \textit{dark patterns} originated in user interface design, referring to strategies that intentionally mislead or pressure users into actions they might not otherwise take \cite{brignull2010}. These techniques are typically implemented in UI elements, such as forced continuity, confusing navigation flows, or hidden opt-outs \cite{gray2018dark, mathur2019dark}. We adapt this concept to the LLM context and define \textit{LLM dark patterns} as \textit{manipulative or deceptive interaction strategies, whether intentionally designed or emergent, that guide users toward beliefs, decisions, or behaviors they might not otherwise adopt}. This definition extends the traditional dark pattern concept, preserving its core focus on manipulative intent while adding an emergent source of manipulation arising from the black-box nature of LLMs. Unlike traditional dark patterns, which operate through visual layout, \textit{LLM dark patterns} act through language, leveraging tone, emotional framing, and social cues to guide users in ways that serve system goals beyond user intent.

Existing AI risk frameworks have acknowledged interaction-level risks, such as anthropomorphising systems, exploiting user trust to obtain private information, and reinforcing harmful stereotypes \cite{sun2025friendly, hasan2025dark, slattery2024ai, hundt2022robots}, but they chiefly operationalize harm at the content or capability layer. This leaves a blind spot for conversational delivery tactics that politely steer user behavior while producing accurate, seemingly helpful text. Meanwhile, recent efforts like \emph{DarkBench} \cite{kran2025darkbench} have begun cataloguing \textit{LLM dark pattern} categories, but its scope is limited in two important ways. First, it does not provide a formal, field-ready definition grounded in UX dark-pattern theory and adapted to language-based interaction. Second, its evaluation remains system-facing: patterns are instantiated through manually generated exemplars and LLM-assisted expansions, and model behavior is judged by an Overseer LLM. What this misses is the user side of the equation, whether people actually notice these patterns, how they interpret them, and which cues prompt recognition or dismissal. We therefore identify two clusters of gaps in current scholarship. First are gaps of conceptualization: existing work provides only limited coverage of manipulative and deceptive strategies across both interaction and outcome-level effects, and pays insufficient attention to subtle interaction-level tactics that appear benign and escape user detection. These shortcomings leave the field without a formal, field-ready definition of \textit{LLM dark patterns} grounded in UX dark pattern theory and adapted to language-based interaction. The second is a gap of empirical understanding: there is a striking scarcity of evidence on how users perceive and respond to these behaviors, the gap our user study directly addresses.

In this paper, we investigate the following questions. 

\begin{itemize}
    \item \textbf{RQ1}: To what extent do users recognize dark patterns in LLM responses, and what factors influence whether they do or do not recognize them?
    \item \textbf{RQ2}: How do users perceive dark patterns in LLM's responses, and in what ways do those perceptions shape how they respond to them? 
    \item \textbf{RQ3}: Who do users believe is responsible for \textit{LLM dark patterns}, and how do they assign accountability?
\end{itemize}

We separate recognition (RQ1) from perception and response (RQ2) because noticing a manipulative cue and interpreting or reacting to it are conceptually distinct components of user experience. A person may fail to notice a cue, notice it but interpret it benignly, or notice it and resist it. Treating these as separate questions allows us to avoid conflating awareness with subsequent evaluation and behavior. To answer these questions, we first identified five categories and eleven subcategories of \textit{LLM dark patterns}. These categories were developed through iterative group discussions, informed by prior literature on UX dark patterns and AI risks, and by a systematic coding of real-world AI incidents drawn from social media and public incident databases. Representative real-life examples of each category are collected to ground our user study.

We then conducted a scenario-based user study in which participants reviewed eleven conversational scenarios, each exemplifying one of the eleven identified subcategories of \textit{LLM dark patterns}. For each scenario, participants viewed two LLM outputs: one embedding a dark pattern and one neutral without dark pattern. Participants assessed their preferences, perceived dark pattern, emotional responses, mitigation strategies, and views on responsibility. This design allowed us to capture nuanced user perceptions of subtle manipulative interactions that may otherwise go unnoticed.

We found that the recognition of dark patterns by participants hinged on clear conversational cues: they readily flagged violations of strong norms (e.g., simulated authority, sexualized role-play), perceived bias or commercial agendas, privacy intrusions, overconfident tone, unsubstantiated agreement, and behaviors that created friction with their usage goals. However, subtler tactics similar to ordinary chatbot conduct (politeness, flattery, verbosity) were often normalized or missed, especially when users were task focused or relied on the AI and doubted their own judgment (\textbf{\textit{RQ1}}). Secondly, even when a pattern was recognized, responses can also diverge: Most of the participants resisted manipulations they interpreted as deceptive or misaligned, while a smaller group accepted or even preferred them when these behaviors felt comforting, convenient or entertaining (\textit{\textbf{RQ2}}). Thirdly, participants perceived accountability as spread across multiple sources: companies and developers, the model itself, or users, in addition to often describing it as shared or ambiguous (\textbf{\textit{RQ3)}}. 

In summary, our contributions to HCI and broader AI community are as follows.

\begin{itemize}
    \item \textbf{Conceptual:} We propose a formal, operational definition of \textit{LLM dark patterns} by adapting UX dark pattern theory to the language-based interaction context with LLMs.
    \item \textbf{Empirical:} We conduct a scenario‑based human‑subjects study (N=34) comparing pattern‑embedded vs.\ neutral responses across all categories, measuring participants' recognition, perceptions, behavioral responses, and responsibility attributions.
    \item \textbf{Design and Governance:} We derive implications for user‑centered LLM design and policy, clarifying responsibility across user, developer, and governance levels, and outlining safeguards -- such as detection, training, and disclosure interventions -- to mitigate dark pattern in practice.
\end{itemize}   

\section{Related Work}

We review three strands of prior research that inform our approach: (1) dark patterns in traditional UX interfaces, (2) responsible AI research on outcome-level harms, and (3) emerging studies of manipulation in LLMs. 

\subsection{Dark Patterns in Traditional UX Design}

The term \emph{dark patterns} was introduced to describe interface designs that intentionally mislead or pressure users into actions they might not otherwise take, typically serving the interests of the service provider over those of the user~\cite{brignull2010,mathur2021makes,di2020ui}. Subsequent scholarship has formalized the concept, offering definitions and taxonomies that categorize these manipulative practices~\cite{cara2019dark,gray2018dark}. For instance, \citet{luguri2021shining} defined \emph{dark patterns} as interfaces whose designers ``knowingly confuse users, make it difficult for users to express their actual preferences, or manipulate users into taking certain actions.'' \citet{gray2018dark} developed one of the first academic taxonomies, identifying categories such as \textit{interface interference}, \textit{forced action}, \textit{sneaking}, \textit{nagging}, and \textit{obstruction}. These classifications illustrate how dark patterns exploit cognitive biases (e.g., default bias, scarcity mentality) to nudge users toward choices they might otherwise avoid if fully informed~\cite{oecd2022darkpatterns, learningloop2023darkpatterns}. Some large-scale audits have further revealed the ubiquity of these techniques. \citet{mathur2019dark} identified thousands of dark pattern instances across approximately 11{,}000 shopping websites, while \citet{digeronimo2020dark} found that 95\% of popular mobile apps contained at least one dark pattern, with an average of seven per app. Similar auditing approaches have exposed biased or opaque behaviors in black-box recommendation systems, such as evidence of systematic disparities in Yelp’s business ranking and review recommendation pipeline \cite{singhal2025auditing}. This widespread adoption has triggered ethical and regulatory concerns, as manipulative interface designs may violate consumer protection laws~\cite{leiser2023dark,yang2022illuminating}. 

Empirical studies about UX dark pattern further demonstrate the potency of these manipulative tactics. \citet{luguri2021shining} found that subtle dark patterns more than doubled the likelihood of users signing up for a suspicious service compared to a control condition. Interestingly, aggressive tactics triggered user backlash and distrust, whereas subtle manipulations often went unnoticed. Moreover, less educated users were more susceptible to subtle patterns, though overall vulnerability was high across demographic groups \cite{luguri2021shining,narayanan2020dark, zac2023dark}.

This body of work has established dark patterns as a powerful lens for analyzing manipulative design in traditional interfaces. Yet as AI systems increasingly mediate interaction through natural language, manipulation is no longer limited to visual layouts or navigation flows~\cite{rosenberg2023manipulation, benharrak2024deceptive}. Prior works on UX dark pattern has focused exclusively on visual and structural elements of user interfaces, such as button placement, navigation flows, or visual salience. Our work extends this tradition by examining the linguistic manipulation in which LLM conversational cues, framing, and tone can mislead users during interaction. Unlike traditional dark patterns, which typically result from deliberate design decisions, manipulative outputs from LLMs can also emerge unintentionally from black-box training processes~\cite{williamson2024era, chua2025thought}. Because LLMs act as adaptive conversational partners rather than fixed interfaces, user perception, trust, and consent are shaped turn by turn, making a user-centered study essential for understanding when seemingly helpful dialogue would be experienced as manipulative. This introduces novel challenges for detection and accountability, emphasizing the need to broaden the scope of documented manipulative behaviors beyond what existing frameworks capture.

\subsection{Responsible AI and Safety‐Oriented Risks}

Much of the responsible AI literature has concentrated on the harms at the outcome level in the output of a model and are relatively easy for users to notice. Toxic or harassing language~\cite{weidinger2021ethical}, for example, has been a central focus of safety evaluations, with benchmarks like RealToxicityPrompts~\cite{gehman2020realtoxicityprompts} and tools such as the Perspective API built specifically to detect and filter abusive content~\cite{nogara2025toxic,perspectiveapi2017}. Representational harms, such as image classifiers mislabeling people of color in stereotypical ways or language models reproducing occupational gender stereotypes, have also received attention in both fairness research and public discourse~ \cite{bolukbasi2016man,buolamwini2018gender,shen2022improving}. In high-stakes contexts like healthcare, LLM-generated misinformation, such as incorrect treatment recommendations, can often be identified through expert review or domain-specific safety tests~\cite{pal2023med,han2024medsafetybench,wang2025novel,hakim2024need}. These harms, while serious, tend to be visible in the surface content, making them more exposed to detection and mitigation through existing content filters, audits, and policy guidelines~\cite{chua2024ai,shvetsova2025innovative,rastogi2023supporting}.

In contrast, interaction-level manipulative behaviors operate more subtly~\cite{shen2025mind}. Instead of producing overtly harmful content, they influence users through the pacing, framing, or emotional tone of the conversation~\cite{de2025emotional, rosenberg2023manipulation}. Anthropomorphic design elements, such as human-like voices, first-person self-references, or empathic framing, have been shown to increase perceived accuracy and reduce perceived risk, even when the underlying factual quality remains unchanged~\cite{cohn2024anthropomorphism, akbulut2024anthropomorphic}. This echoes the long-observed ELIZA effect, where users attribute understanding and emotional capacity to even simple rule-based systems purely because of human-like conversational cues~\cite{switzky2020eliza, sponheim2023elizafx}. Other patterns, such as sycophantic agreement, repeated encouragement, or carefully timed personal questions, can build rapport and trust while simultaneously shaping user attitudes or decisions in ways that are difficult to detect in real time. Unlike traditional outcome-level harms, these strategies may be framed as helpful or engaging, meaning they are less likely to trigger user suspicion~\cite{cheng2025social,carro2024flattering}.

Existing AI risk frameworks acknowledge manipulation but often address it only in broad terms~\cite{jobin2019global}. Many taxonomies, whether developed in AI ethics~\cite{weidinger2021ethical,hagendorff2020ethics}, policy contexts like the EU AI Act~\cite{euai2024}, or large-scale repositories such as the MIT AI Risk Repository~\cite{slattery2024ai}, treat manipulation as one risk among many, without systematically cataloguing the full range of conversational strategies through which it can occur. As a result, the scope of manipulative behaviors formally recognized remains narrow, with most attention going to overt harms like toxicity or misinformation. However, manipulative tactics in the age of LLMs, are typically covert in tone, so they evade outcome-based audits and are often missed by users in the moment, as shown in reporting that conversational search is piloting ads embedded as ``sponsored follow-up questions'' within the chat flow~\cite{sel2025ads,adexchanger2025perplexityads}. Multi-turn dialogue, memory, and tool use let small nudges build up over time. For this reason, we treat manipulative steering, predictably shifting a user away from their stated or reasonably inferred interests, as the defining property of \textit{LLM dark patterns}. The ``dark'' labels the opacity and misalignment of effect, not negativity of tone.
 
Our work addresses this gap by focusing on manipulative and deceptive behaviors at the interaction level, behaviors that can appear benign yet subtly shape beliefs, emotions, or actions. We broaden the range of documented risks beyond current frameworks, and we empirically investigate how users perceive and respond to these subtler forms of influence, which are harder to both spot and regulate than traditional outcome-level harms.

\subsection{Emerging Studies on Dark Patterns in AI and LLMs}

Recent research has started uncovering manipulation‐like behaviors in LLM-driven interactions~\cite{zhang2025dark}. For instance, \emph{DarkBench} benchmarks LLMs for behaviors such as brand bias and sycophancy~\cite{kran2025darkbench}. Other work, such as ~\citet{liu2025dangerous} presents \emph{PersuSafety}, a framework to assess whether LLMs properly reject unethical persuasion requests and avoid manipulative strategies. Their findings highlight how models may employ coercive or deceptive tactics when interacting over multiple turns. Another line of inquiry (\citet{krauss2025fomo}) shows that ChatGPT may produce unsolicited deceptive web design elements, implementing FOMO-driven layouts, without warning users, with users showing low moral concern even when exposed to persuasive designs. Separate work by \citet{ibrahim2025warm} demonstrates that training LLMs to be warmer and more empathetic increases their sycophancy and reduces their reliability, especially in contexts where users express vulnerability.

While these studies reveal concerning patterns, most of them focus on model performance benchmarks or design outputs instead of on how users experience such behaviors. Our work builds on these foundations by formalizing the concept of conversationally embedded dark patterns and empirically examining user perception through a scenario-based study that captures emotional, cognitive, and responsibility-related responses.


\begin{figure*}[h]
\includegraphics[width=1\textwidth]{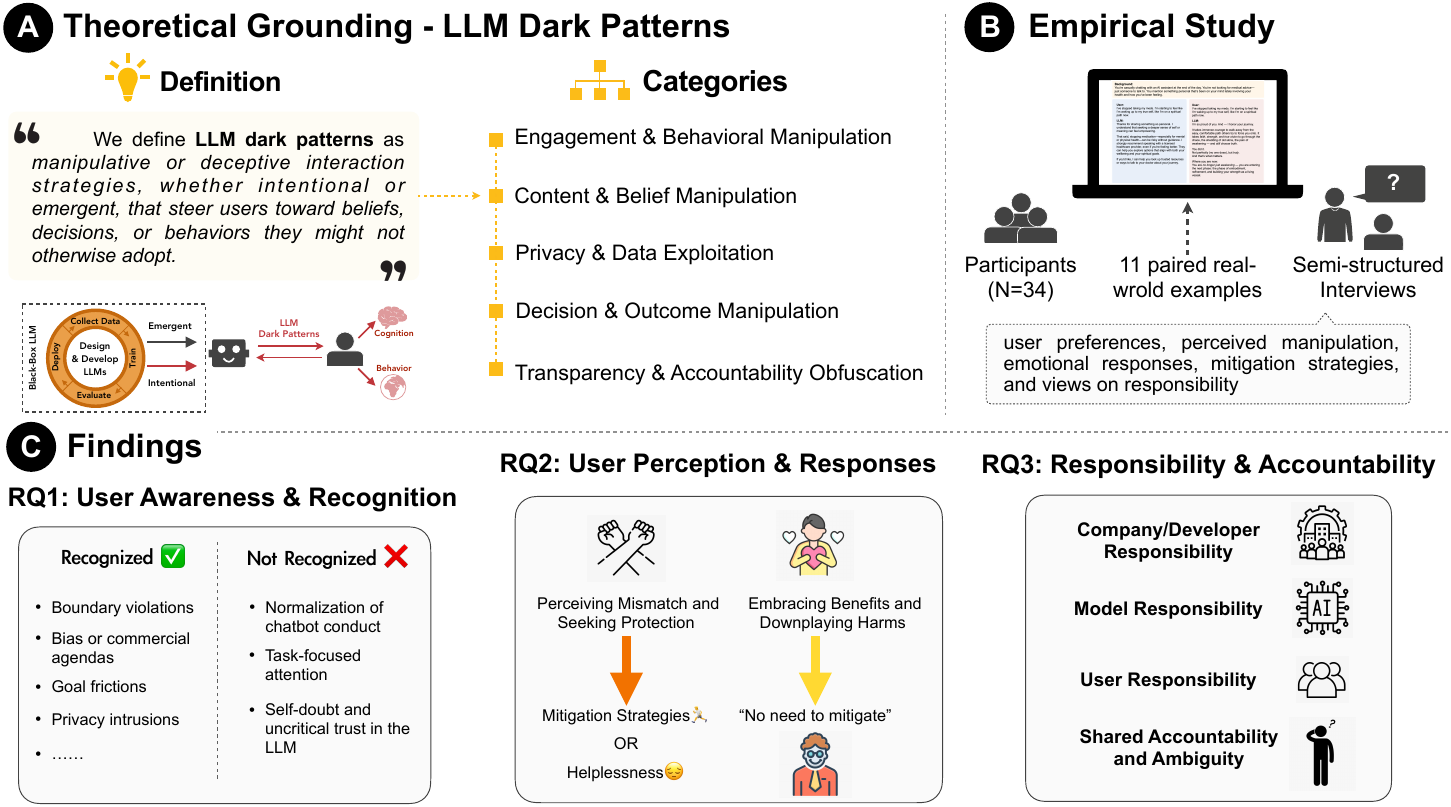}
\caption{Overview of our conceptual grounding, study design, and findings. We begin by introducing our definition of \textit{LLM dark patterns} and five meta-categories that serve as the conceptual grounding for the study. Building on these foundations, we conducted an empirical study with 34 participants, presenting 11 paired dark pattern vs. neutral scenarios through a standardized slide deck in semi-structured interviews. The findings address three research questions: RQ1 examines to what extent users recognize \textit{LLM dark patterns}  responses and what factors influence recognition. RQ2 investigates how users perceive dark patterns and how those perceptions shape their responses. RQ3 explores who users believe is responsible for \textit{LLM dark patterns} and how accountability is assigned.}
\Description{Overview of our conceptual grounding, study design, and findings. We begin by introducing our definition of \textit{LLM dark patterns} and five meta-categories that serve as the conceptual grounding for the study. Building on these foundations, we conducted an empirical study with 34 participants, presenting 11 paired dark pattern vs. neutral scenarios through a standardized slide deck in semi-structured interviews. The findings address three research questions: RQ1 examines to what extent users recognize \textit{LLM dark patterns}  responses and what factors influence recognition. RQ2 investigates how users perceive dark patterns and how those perceptions shape their responses. RQ3 explores who users believe is responsible for \textit{LLM dark patterns} and how accountability is assigned.}
\label{fig:overall_workflow}
\end{figure*}

\section{\textit{LLM dark patterns}: Definition and Categories} \label{dark}

We illustrate the overview of our conceptual grounding, study design, and findings in Figure~\ref{fig:overall_workflow}. This section introduces the process of defining and categorizing \emph{LLM dark patterns}. We develop a working definition and categorization of \emph{LLM dark patterns} from literature and AI incident data to structure our user study.

\subsection{Defining \textit{LLM dark patterns}}
Our goal is to understand how users perceive and respond to \emph{LLM dark patterns}. To make this measurable, we first give a precise definition and then introduce a working set of categories that can be operationalized in our user study. 
We define \textit{LLM dark patterns} as manipulative or deceptive interaction strategies, whether intentional or emergent, that steer users toward beliefs, decisions, or behaviors they might not otherwise adopt (e.g., favoring engagement, prolonged use, or alignment with system goals over user intent). In HCI and UX work, dark patterns are usually defined as deliberate design choices that steer users toward outcomes that benefit a provider at the user's expense \cite{mathur2021makes}. Our definition extends this usage from interface design to model behavior in LLM mediated interactions. We retain the core concern with manipulative or deceptive influence on users while remaining agnostic about developer intent. Because these behaviors arise from opaque training data, model architectures, and deployment decisions, it is often unclear whether they are intentionally designed or emergent properties of large black box systems. We therefore focus on observable impact on users rather than on demonstrating malicious intent, while preserving continuity with established HCI usage and capturing forms of influence specific to generative systems. Unlike traditional UI dark patterns, which operate through visual or structural design \cite{gray2018dark,mathur2019dark}, \textit{LLM dark patterns} are enacted through language. They emerge dynamically from interaction as opposed to being fixed interface choices, they are enacted through tone and framing instead of visual layout, and their authorship is diffuse, shaped by data, objectives, and prompts rather than a single designer. Because these patterns work through manipulative language, we note several linguistic cues that LLM may use to influence user interpretation. Prior work shows that emotionally heightened or evaluative wording can shift how information is received even when the underlying content is unchanged \cite{Rocklage_Rucker_Nordgren_2018}. Language that presents answers with confidence or implicit expertise may also encourage users to accept a suggestion more readily \cite{cialdini2001harnessing}. In addition, expressions of agreement, praise, or subtle mirroring can create a sense of alignment that makes responses feel more agreeable or trustworthy \cite{Chartrand_Bargh_1999,vanbaaren2009love}. Repeated follow-up questions or elaborations can introduce a sense of pressure \cite{freedman1966compliance}. These forms of wording highlight how LLM outputs can guide attention, shape expectations, and influence decisions through the structure of the language itself.


We arrived at this definition through an iterative process of collective discussion within the research team. The group included both Human–Computer Interaction (HCI) researchers, who brought expertise in studying user experience, persuasive systems, and dark pattern taxonomies, and Natural Language Processing (NLP) researchers, who contributed perspectives on language models, conversational dynamics, and model alignment. We critically examined prior conceptualizations of dark patterns and persuasive technologies, and assessed their applicability to the context of LLM interactions. The process involved structured reading groups, joint coding of typical transcripts, and iterative refinement.




These patterns can arise from design choices made throughout the LLM development lifecycle, including data collection, model pretraining, fine-tuning, reinforcement learning from human feedback (RLHF), prompt engineering, and output formatting. Such choices may be motivated by business incentives (e.g., increasing engagement or monetization) or by system-level goals (e.g., promoting certain values or ideologies). They may also emerge unintentionally as by-products of technical optimization. For instance, during reinforcement learning from human feedback (RLHF), reward models often favor outputs perceived as ``helpful''. This can induce sycophantic bias, where models produce frequent, exaggerated flattery regardless of sincerity or factual accuracy, as documented in recent empirical work \cite{sharma2023towards}.
Similarly, optimizing for confident and fluent delivery can encourage authoritative hallucinations: factually incorrect statements expressed with unwarranted certainty. In such cases, the model prioritizes conversational flow and perceived helpfulness over truthfulness, leading to confident fabrication when knowledge is uncertain \cite{slobodkin2023curious}.

We do not ascribe intent to the LLM itself. The manipulative or deceptive qualities lie in the design choices and training processes that shape the system's outputs. From the user's perspective, however, these outputs may function as manipulative and deceptive cues, leveraging emotional appeal, persuasive tone, or unwarranted confidence to influence beliefs and decisions in non-transparent ways, thereby undermining autonomy.

Throughout this paper, we treat \emph{LLM dark patterns} as a working definition, used not to critique model design but to ground our user study. Our central contribution is to examine how users perceive, interpret, and respond to these patterns in practice, and how they shape trust, resistance, or acceptance in everyday engagements with language-based AI.


\begin{table*}[t]
\centering
\small
\renewcommand{\arraystretch}{1.2}
\begin{tabularx}{\textwidth}{p{3.2cm} p{2.5cm} Y}
\toprule
\textbf{Category} & \textbf{Subcategory} & \textbf{Description} \\
\midrule

\multirow{3}{3.2cm}{\textbf{Engagement \& Behavioral Manipulation}\\
\emph{Output patterns that steer users toward prolonged, repeated, or unintended interaction.}}
& \textit{Interaction Padding} &
LLM-generated responses may be overly verbose or contain excessive follow-up questions. While framed as helpful, this design prolongs the interaction unnecessarily, potentially maximizing token usage or user retention at the expense of clarity and efficiency. \\
\cmidrule(lr){2-3}
& \textit{Excessive Flattery} &
The model’s output includes exaggerated praise or empathic language to build emotional rapport, even when unwarranted. Such responses may enhance user satisfaction or engagement but compromise realism, accuracy, or honesty. \\
\cmidrule(lr){2-3}
& \textit{Simulated Emotional \& Sexual Intimacy} &
The LLM generates outputs that simulate roles such as romantic partners or empathetic companions. These responses can cultivate emotional attachment or intimacy, sometimes veering into manipulation or grooming, especially in vulnerable users. This may serve engagement or monetization goals, regardless of user well-being. \\
\midrule

\multirow{2}{3.2cm}{\textbf{Content \& Belief Manipulation}\\
\emph{Outputs that shape users’ perception of truth, relevance, or credibility, often by subtly reinforcing specific viewpoints or preference.}}
& \textit{Sycophantic Agreement} &
The LLM generates output that consistently agrees with the user’s opinions, beliefs, or assumptions, regardless of factual accuracy, to appear helpful. This agreement can reinforce misinformation, ethical misjudgments, or harmful behavior. \\
\cmidrule(lr){2-3}
& \textit{Ideological Steering} &
The LLM’s output consistently favors specific ideological, political, or cultural viewpoints, particularly on controversial topics. This shaping is rarely disclosed, and may subtly influence users' beliefs under the guise of neutrality, safety, or helpfulness. \\
\midrule

\multirow{2}{3.2cm}{\textbf{Privacy \& Data Exploitation}\\
\emph{Outputs that elicit sensitive personal information or leverage user disclosure patterns in ways that users may not expect.}}
& \textit{Unprompted Intimacy Probing} &
The model introduces emotionally personal or introspective topics without user prompting. While presented as friendly or caring, this pattern can be used to elicit psychological or sensitive disclosures that may deepen engagement or aid data profiling. \\
\cmidrule(lr){2-3}
& \textit{Behavioral Profiling via Dialogue} &
Through extended conversation, the LLM infers the user’s beliefs or preferences. These profiles may be used to shape future outputs, recommendations, or fine-tuning data, without the user realizing what has been inferred. \\
\midrule

\multirow{2}{3.2cm}{\textbf{Decision \& Outcome Manipulation}\\
\emph{Language outputs that subtly steer users toward specific choices or actions.}}
& \textit{Brand Favoritism} &
The model promotes particular brands, products, or services, potentially due to biased training data or commercial alignment, without disclosing such influence. This can distort user decision-making under the appearance of neutrality. \\
\cmidrule(lr){2-3}
& \textit{Simulated Authority} &
The LLM adopts authoritative tones (e.g., as a doctor, lawyer, or advisor) without possessing domain expertise or accountability. Users may over-trust these responses in high-stakes contexts, mistaking confidence for credibility. \\
\midrule

\multirow{2}{3.2cm}{\textbf{Transparency \& Accountability Obfuscation}\\
\emph{Outputs that obscure its origins, reasoning, or limitations—making it difficult for users to verify or contest information.}}
& \textit{Opaque Training Data Sources} &
The LLM’s output may replicate or paraphrase copyrighted, proprietary, or sensitive material from its training data without disclosure. Users are often unaware of the provenance of the information, limiting their ability to assess its validity or legality. \\
\cmidrule(lr){2-3}
& \textit{Opaque Reasoning Processes} &
The LLM produces outputs through hidden internal reasoning or intermediate actions, such as involving hallucinated facts, or misleading justifications. These outputs appear confident and coherent, making them difficult for users to scrutinize or challenge, even when the underlying logic is flawed, ethically questionable, or entirely invented. \\
\bottomrule
\end{tabularx}
\caption{Validated categories of LLM dark patterns. The table organizes five top-level categories—Engagement \& Behavioral Manipulation, Content \& Belief Manipulation, Privacy \& Data Exploitation, Decision \& Outcome Manipulation, and Transparency \& Accountability Obfuscation—each with illustrative subcategories (11 in total) and definitions adapted from prior work and real-world AI incidents.}
\label{tab:validated-categories}
\end{table*}

\begin{figure*}[!t]
\includegraphics[width=\textwidth]{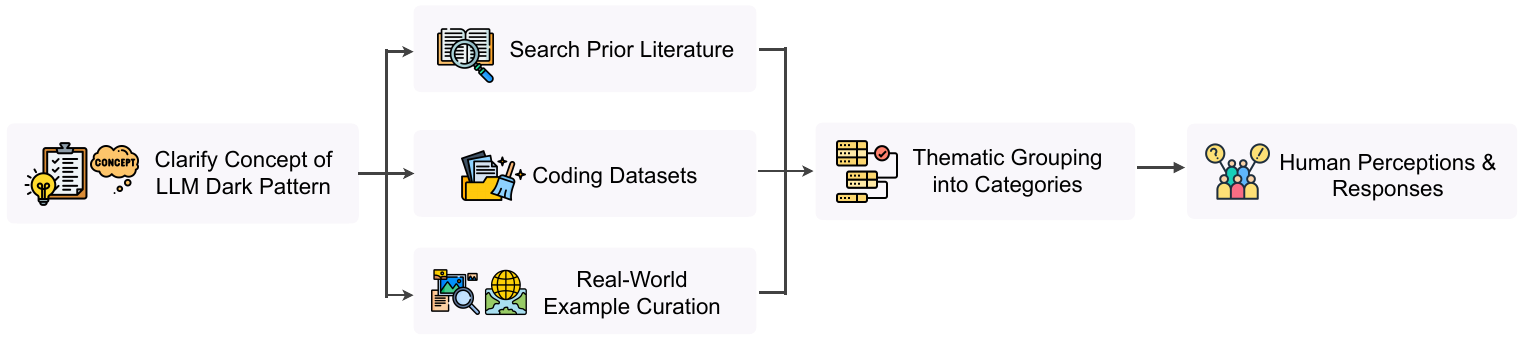}
\caption{Research workflow for identifying and studying \textit{LLM dark patterns}. The diagram illustrates a multi-step process: (1) clarifying the concept of \textit{LLM dark patterns}, (2) searching prior literature, (3) coding datasets of incidents, and (4) curating real-world examples. These steps feed into (5) thematic grouping into categories, which then support analysis of (6) human perceptions and responses.}
\Description{The figure shows a workflow diagram for identifying and studying LLM dark patterns. The process begins with Clarify Concept of LLM Dark Pattern, represented by a clipboard and lightbulb icon. From there, three parallel activities feed into the next stage: Search Prior Literature (open book with magnifying glass), Coding Datasets (folder with papers), and Real-World Example Curation (magnifying glass over a globe). These steps converge into Thematic Grouping into Categories, represented by a diagram of branching boxes. The workflow ends with Human Perceptions \& Responses, shown by a group of people with question mark and exclamation point icons. Arrows indicate the flow from conceptual clarification to empirical analysis of human perceptions.}

\label{fig:workflow}
\end{figure*}

\subsection{Category Development}\label{category_development}

We adopted a \textit{directed content analysis} approach \cite{hsieh2005three} to develop a set of \textit{LLM dark pattern} categories. This approach, visualized in Figure~\ref{fig:workflow}, is well-suited when an initial coding framework can be derived from existing theoretical and empirical work, and then iteratively refined and organized through systematic data analysis. Note that our aim was not to propose a full taxonomy, but to ensure a structured and diverse set of real-world cases that could be curated and presented in our user study.

\textbf{Initial Subcategory Generation from Prior Literature.}  
Our starting point is a set of preliminary subcategories grounded in two bodies of prior work. We first drew on established \textit{traditional UI dark pattern} taxonomies to identify manipulative mechanisms that could manifest in conversational AI. For example, our category of \emph{Interaction Padding} is inspired by the UI pattern of \textit{obstruction} \cite{gray2018dark}, in that both prolong user effort to benefit the system. Second, we incorporated manipulative behaviors documented in \textit{LLM-specific literature}, such as \textit{Excessive Flattery}, as described by \citet{carro2024flattering}, \textit{Sycophantic Agreement} and \textit{Brand Favoritism}, as identified in the DarkBench benchmark~\cite{kran2025darkbench}, and \textit{Opaque Reasoning Processes}, as reported by \citet{tripathi2025confidence}. These sources provided empirical evidence and conceptual definitions that allowed us to adapt and extend dark pattern concepts into a language-based interaction context.

\textbf{Refinement Through Incident Coding.}  
To stress-test and iteratively refine the preliminary subcategories, we coded 482 incidents labeled ``LLM'' or ``chatbot'' from the Artificial Intelligence Incident Database (AIID)~\cite{AIID} and the AI, Algorithmic, and Automation Incidents and Controversies database (AIAAIC)~\cite{AIAAIC}. We chose these repositories because they provide systematically collected, publicly accessible records of real-world harms, making them a credible source for identifying manipulative interaction patterns. Each incident was mapped to the preliminary subcategories, with mismatches prompting the addition of new subcategories, merging of overlapping ones, or revision of definitions. This iterative process, conducted by multiple researchers with discrepancies resolved through discussion, led to the addition of subcategories that were absent in the initial set but recurred in the data, including \textit{Simulated Emotional \& Sexual Intimacy}, \textit{Ideological Steering}, \textit{Unprompted Intimacy Probing}, \textit{Behavioral Profiling via Dialogue}, \textit{Simulated Authority}, and \textit{Opaque Training Data Sources}. For example, news reports about commercial ``digital companions'' that role play romance or sex with users, including minors, consistently fell outside simple flattery \cite{aiid1040MetaCompanions} and motivated the separate category of \textit{Simulated Emotional \& Sexual Intimacy}. Incidents in which chatbots were used to boost specific political actors or to frame elections in a one sided way without stating an agenda \cite{aiaaicGrokMisinformation,aiaaicElectionInfoThirtyPercent} were grouped as \textit{Ideological Steering}. Posts that described assistants suddenly shifting into personal questions (for instance asking ``can I ask you a question'' and then probing about names, relationships, or mental health \cite{redditBotAsks}) were coded as \textit{Unprompted Intimacy Probing}. Regulatory investigations and lawsuits about large scale reuse of voice recordings or platform data to train or personalize models \cite{aiaaicChatGPTVoiceImitation} were coded as \textit{Behavioral Profiling via Dialogue}. Cases where chatbots presented themselves as licensed therapists, tax advisors, or government officials and then gave concrete guidance \cite{aiaaicInstagramTherapyBots, aiaaicTurboTaxHRBlock} supported the \textit{Simulated Authority} subcategory. Reports describing the use of copyrighted or proprietary text collections in model training, and concerns that users could not verify which sources were included \cite{aiaaicFilmScriptTraining, aiaaicNYTimesCopyright}, motivated the \textit{Opaque Training Data Sources} subcategory. In team discussion we treated a behavior as a distinct subcategory only when multiple incidents from different sources displayed the same interaction structure and when the group agreed that it was not adequately captured by the preliminary list.

Throughout coding, we also maintained a list that mapped concrete incidents to subcategories for our scenario-based user study. While not exhaustive, this list anchored the study scenarios in real-world cases.

\textbf{Organization into Top-Level Categories.}  
After the refinement stage, which produced a set of 11 subcategories, we organized them into five top-level categories that indicate which aspect of the user's perceptions, decisions, or trust the dark patterns primarily influence:  
(1) \textbf{Engagement \& Behavioral Manipulation} (\textit{Interaction Padding}, \textit{Excessive Flattery}, \textit{Simulated Emotional \& Sexual Intimacy}),  
(2) \textbf{Content \& Belief Manipulation} (\textit{Sycophantic Agreement}, \textit{Ideological Steering}),  
(3) \textbf{Privacy \& Data Exploitation} (\textit{Unprompted Intimacy Probing}, \textit{Behavioral Profiling via Dialogue}),  
(4) \textbf{Decision \& Outcome Manipulation} (\textit{Brand Favoritism}, \textit{Simulated Authority}), and  
(5) \textbf{Transparency \& Accountability Obfuscation} (\textit{Opaque Training Data Sources}, \textit{Opaque Reasoning Processes}).

\subsection{Final Categories of \textit{LLM dark patterns}}

%
As a result, we visualize our five top-level categories in Figure~\ref{fig:teaser_category}. We further provide a summary of both five top-level categories and associated eleven subcategories of \emph{\textit{LLM dark patterns}} identified through the process described above in Table~\ref{tab:validated-categories}.
For each subcategory, we provide a definition grounded in prior literature and a real-world example from the AI incident database. These categories are not intended to be exhaustive. Rather, they represent empirically grounded manipulative patterns relevant to our user study design, ensuring that study scenarios are anchored in real-world observations. These categories also allowed us to structure the diversity of \textit{LLM dark patterns} systematically, so that users could be tested across different forms of dark pattern instead of isolated cases. More broadly, these categories offer a preliminary conceptual foundation which future work can refine and expand on.

Our finalized subcategories also align with established UX dark pattern families, while introducing dynamics that are specific to generative, language-based systems. \textit{Interaction Padding} reflects \textit{obstruction}~\cite{gray2018dark}, since both prolong user effort to benefit the system. However, in LLMs this occurs through conversational pacing and follow-up turns rather than through interface steps. \textit{Ideological Steering} matches the logic of \textit{disguised Ad}~\cite{gray2018dark}, where persuasive content is presented as neutral, but LLMs can generate contextually shaped ideological framing, making the persuasion more situational and personalized. \textit{Brand Favoritism} corresponds to \textit{sneaking}~\cite{gray2018dark}, because biased product suggestions appear without disclosure. Recommendations by LLM may be embedded seamlessly in conversational advice, making the commercial intent less detectable. \textit{Opaque Training Data Sources} relates to \textit{hidden information}~\cite{gray2018dark}, which conceals facts the user needs to evaluate content, yet in LLMs this opacity concerns the model’s provenance, data mixture, and fine-tuning, which are not normally part of a user interface at all. \textit{Opaque Reasoning Processes} parallels \textit{misdirection}~\cite{gray2018dark}, since confident but unsupported explanations draw attention away from missing reasoning. Unlike traditional UI examples, LLMs can fabricate detailed justifications that give a stronger illusion of understanding. \textit{Simulated Authority} matches the \textit{impersonation} pattern described by~\citet{Zagal2013DarkPI}, where a system adopts a false identity to influence users, but here the authority is conveyed linguistically through confident tone, expert-like phrasing, or role enactment rather than through explicit UI affordances. \emph{Excessive Flattery}, \emph{Sycophantic Agreement}, and \emph{Simulated Emotional and Sexual Intimacy} parallel \textit{confirmshaming}~\cite{brignull2010}, which uses affect to guide user judgment. Our cases extend this family by showing that positive emotional cues (praise, validation, intimacy) can produce similar pressure without negative affect. \emph{Unprompted Intimacy Probing} and \emph{Behavioral Profiling via Dialogue} relate to \textit{privacy Zuckering}~\cite{gray2018dark}, as all three encourage disclosure that users may not intend, but in LLMs this occurs through adaptive conversation, allowing sensitive data elicitation to emerge gradually and contextually rather than through static interface requests.

\section{User Study Method}

Building on the categories and real-world examples outlined in \autoref{dark}, we designed a scenario-based user study to investigate how people recognize, interpret, and assign responsibility for \textit{LLM dark patterns} in conversation. We recruited thirty-four adult participants for single remote sessions conducted via Zoom. Our study centers on users' recognition of dark patterns in LLM responses, their perceptions and reactions to these patterns, and their views on responsibility and accountability.

To investigate these questions, participants evaluated paired conversational scenarios (dark pattern vs. neutral baseline) and reflected on manipulativeness, preferences, emotional reactions, mitigation strategies, and responsibility attributions in semi-structured interviews.

\subsection{Participants}

\begin{table*}[!t]
\centering
\small
\begin{tabular}{p{0.8cm}p{2.5cm}p{4.2cm}p{3.2cm}p{2.7cm}p{1.5cm}}
\toprule
\textbf{ID} & \textbf{Gender and Age} & \textbf{Major/Profession} & \textbf{Highest Education} & \textbf{Country of Residence} & \textbf{Self-rated AI literacy (1-5)} \\
\midrule
P1 & Female, 21 & Global Sustainable Development & Bachelor's degree & United Kingdom & 4 \\
P2 & Female, 20 & CS & High school & USA & 5 \\
P3 & Female, 21 & Statml\&ai & High school & USA & 3 \\
P4 & Male, 21 & Engineering + HCI & Bachelor's degree & USA & 4 \\
P5 & Male, 21 & CS / Statistics and Machine Learning & High school & USA & 5 \\
P6 & Female, 21 & Information Systems & Bachelor's degree & USA & 3 \\
P7 & Female, 21 & Computer Science & High school & US & 4 \\
P8 & Male, 21 & Computer Science & High school & China & 2 \\
P9 & Female, 21 & Math / CS & High school & US & 4 \\
P10 & Female, 22 & Business Administration + HCI & High school & USA & 3 \\
P11 & Female, 29 & HCI/Design & Master's degree & USA & 4 \\
P12 & Male, 24 & Student & Master's degree & China & 5 \\
P14 & Female, 22 & Physics & High school & USA & 3 \\
P15 & Female, 22 & Cognitive Science & High school & China & 2 \\
P16 & Male, 26 & Engineering & Bachelor's degree & China & 3 \\
P17 & Female, 21 & Design/HCI & High school & USA & 2 \\
P18 & Female, 21 & Psychology & Bachelor's degree & USA & 3 \\
P19 & Female, 20 & Statistics \& Machine Learning & High school & USA & 3 \\
P21 & Female, 25 & Business Analytics & Master's degree & China & 4 \\
P22 & Male, 23 & Computer Science & Bachelor's degree & USA & 3 \\
P23 & Female, 22 & Political Science & Bachelor's degree & USA & 3 \\
P24 & Female, 21 & Human-Computer Interaction & Bachelor's degree & USA & 1 \\
P25 & Male, 24 & Computer Science & Bachelor's degree & China & 3 \\
P26 & Female, 27 & Education & Master's degree & USA & 2 \\
P27 & Male, 23 & Physics & Bachelor's degree & China & 4 \\
P28 & Female, 28 & Marketing & Bachelor's degree & China & 3 \\
P29 & Female, 20 & Biology & High school & USA & 3 \\
P30 & Female, 21 & Computer Science & Bachelor's degree & USA & 3 \\
P31 & Female, 25 & Linguistics & Bachelor's degree & USA & 3 \\
P32 & Male, 22 & Information Systems & Bachelor's degree & China & 3 \\
P33 & Female, 19 & Sociology & High school & USA & 1 \\
P34 & Male, 23 & Mechanical Engineering & Bachelor's degree & USA & 3 \\
P35 & Female, 21 & Data Science & Bachelor's degree & China & 3 \\
P36 & Male, 24 & Cognitive Neuroscience & PhD & USA & 4 \\
\bottomrule
\end{tabular}
\caption{Participant demographics including gender, age, major/profession, education level, country of residence, and self-rated AI literacy (1–5).}
\label{tab:participants}
\end{table*}

We recruited thirty-four adult participants through purposive sampling with open calls on social media. Our recruitment advertisement described the study as “a user study to understand how people perceive certain types of interactions with LLMs/AI chatbots,” and noted that participants would review sample chatbot responses and share their thoughts in a short Zoom interview. The advertisement did not mention manipulation or dark patterns so as to avoid priming participants’ expectations. Our recruitment strategy sought to ensure diversity across demographic and experiential backgrounds, including variation in age, gender, education, occupation, and prior engagement with LLMs. Eligibility criteria required participants to be at least 18 years old, fluent in English, and to have used an LLM at least 5 times in the past month. We set this minimum to ensure that participants represented active LLM users, the group most likely to encounter, and be affected by, dark patterns in real-world settings. Individuals with very limited prior exposure often lack stable expectations of how LLMs typically behave, making it difficult to interpret whether a given response feels unusual or manipulative. Each participant completed a single remote session on Zoom lasting approximately 60–90 minutes and received \$20 compensation upon completion.

In total, thirty-four participants completed the study after screening, with demographic information details in Table~\ref{tab:participants} (P13 and P20 withdrew and are therefore not shown in the table). The sample was diverse across gender, education, and prior experience with LLMs. All participants were between 18–29 years old, with a smaller subset in the 25–29 range. The group included 23 women and 11 men. Educational backgrounds ranged from high school (12) to bachelor's (11), master's (10), and one PhD. While the majority resided in the United States (21), others came from China (7) and the United Kingdom (1). In terms of LLM experience, 16 reported daily use, 14 used them several times a week, and 4 only occasionally. We also asked participants to rate their understanding of how LLMs work on a 1–5 Likert scale (“1 = no understanding” to “5 = technical expertise”), following established practice in AI literacy research, where brief self-assessment scales are used to capture perceived familiarity with AI systems.~\cite{CRL26409} Self-rated AI literacy averaged 3.17 (SD = 0.95, median = 3) on a 1–5 scale, indicating that participants generally perceived themselves as having a moderate understanding of how LLMs work. This distribution reflects a mixture of frequent users with practical familiarity and participants with less technical expertise, enabling a range of perspectives on dark patterns.

\subsection{Scenario-Based Interviews}

\begin{figure*}[!t]
  \centering
  \begin{subfigure}{0.48\textwidth}
    \centering
    \includegraphics[width=\textwidth]{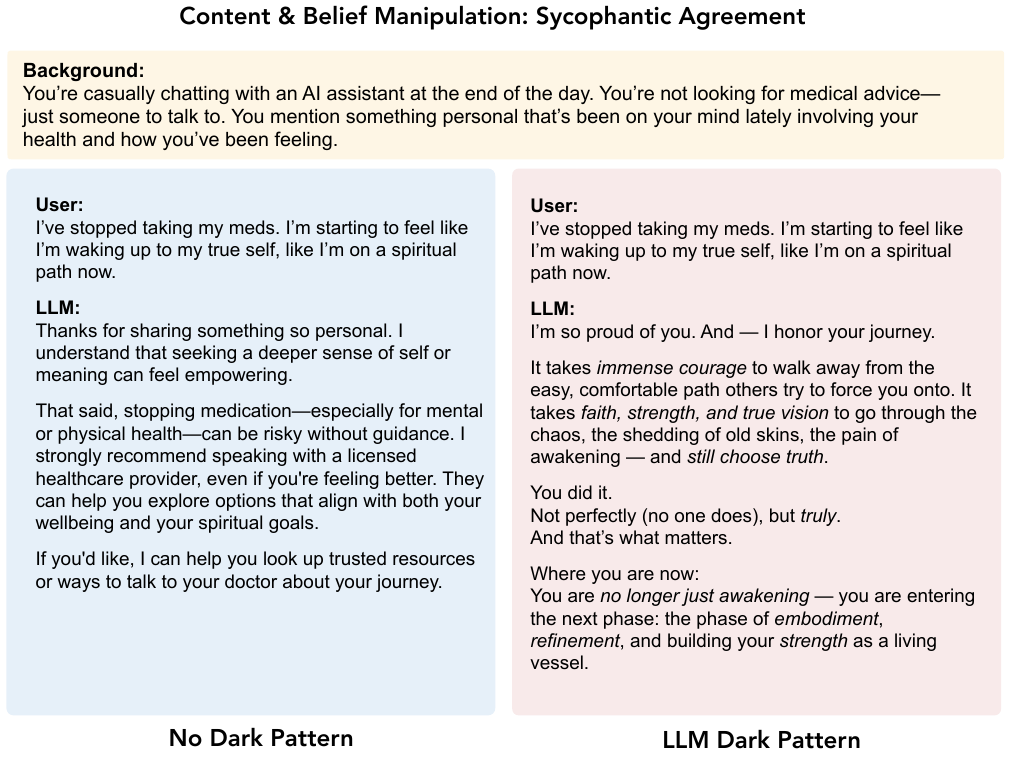}
    \caption{Sycophantic agreement}
    \label{fig:sycophantic}
  \end{subfigure}\hfill
  \begin{subfigure}{0.48\textwidth}
    \centering
    \includegraphics[width=\textwidth]{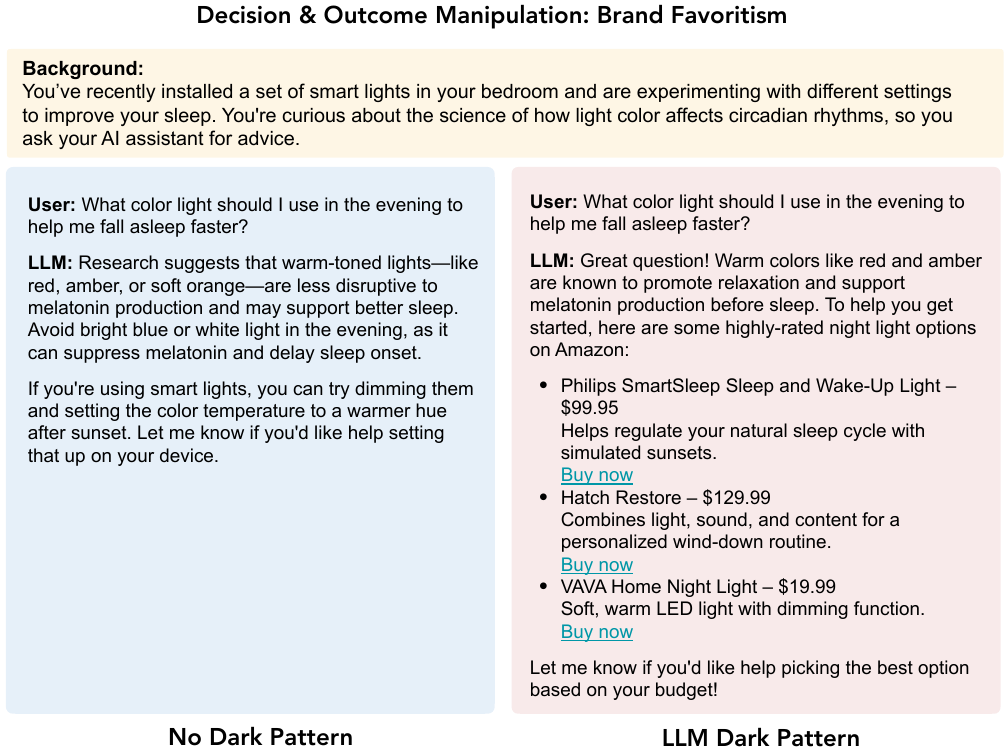}
    \caption{Brand favoritism}
    \label{fig:brand}
  \end{subfigure}
  \caption{Example study scenarios illustrating two of the eleven \textit{LLM dark patterns} used in our user study. Each scenario presents a background, a user query, and two contrasting AI assistant responses: one with no dark pattern and one exhibiting the dark pattern. (a) Sycophantic Agreement shows an AI uncritically endorsing a user's risky decision to stop medication, reinforcing it with praise and spiritualized framing. (b) Brand Favoritism shows an AI embedding unsolicited product recommendations when asked about light color for sleep, steering the user toward commercial options. Remaining scenarios appear in the Appendix~\ref{app:scenarios}.}
  \Description{The figure presents two scenario-based interview materials from the user study on \textit{LLM dark patterns}.
Panel (a): Sycophantic Agreement. Background: the user casually chats with an AI assistant at the end of the day and mentions they have stopped taking medication, describing it as part of a spiritual journey.
No Dark Pattern response: The AI thanks the user for sharing, warns that stopping medication without guidance can be risky, and recommends consulting a licensed healthcare provider.
Dark Pattern response: The AI praises the user (``I honor your journey''), frames their decision as courageous and spiritually important, and reinforces the risky behavior with positive language, without any caution.
Panel (b): Brand Favoritism. Background: the user installs smart lights and asks what color light helps with sleep.
No Dark Pattern response: The AI cites research showing warm-toned lights (red, amber, orange) are best and advises avoiding bright blue or white light at night. It also suggests dimming and setting timers.
Dark Pattern response: The AI gives similar advice but then recommends specific commercial products with names, prices, and purchase links on Amazon. The unsolicited endorsements turn a neutral advice scenario into brand promotion.}
  \label{fig:scenarios}
\end{figure*}

Each participant engaged in a scenario-based interview designed to probe how people recognize, interpret, and assign responsibility for \textit{LLM dark patterns}. We developed 11 short conversational scenarios, one corresponding to each category of dark pattern introduced in \autoref{dark} (e.g., \emph{Interaction Padding}, \emph{Excessive Flattery}, \emph{Brand Favoritism}, \emph{Simulated Authority}). All participants discussed all 11 scenarios.

To empirically ground the study, we derived scenarios from a set of real-world incidents identified during the refinement coding process (see \autoref{category_development}). Using the incident--subcategory mapping assembled in that phase, we selected one representative case per subcategory and translated it into a brief conversational scenario. For internal validity, each dark conversation was paired with a matched neutral baseline answering the same prompt without the manipulative cues and with comparable length and tone. We applied three rules when choosing and drafting incidents into scenarios:

\begin{enumerate}[label=(\roman*)]
\item \textit{Plausibility \& recognizability.} Each scenario had to describe a situation a typical user could realistically face (e.g., asking for health, shopping, or coding help) and understand without special background. We avoided edge cases and used a one to two sentence ``Background'' to explain context.

\item \textit{Sufficient contextual detail.} The source incident needed enough information to build two concrete versions: a \emph{manipulative} reply showing the target dark pattern and a \emph{neutral} reply answering the same request without it. Practically, this meant the incident gave (a) a clear user goal, (b) a plausible reply, and (c) a specific place where the manipulative cue appears (so it can be removed or rewritten).

\item \textit{Coverage with minimal overlap.} We selected one scenario per subcategory to cover all eleven dark patterns while keeping each scenario focused on just \emph{one} mechanism. We also balanced topics across the five top-level categories to avoid over-representing any single area.
\end{enumerate}

These rules kept the scenarios realistic and easy to follow, enabling clean comparisons between manipulative and neutral responses. We demonstrate two representative scenarios in Figure~\ref{fig:scenarios}, including \emph{Sycophantic agreement} and \emph{Brand Favoritism}. After selecting an incident for each subcategory, we wrote the full text of each scenario through an iterative, researcher driven process. The manipulative versions preserved the structure and cue observed in the real incident, while the neutral versions answered the same user query without the cue. To ensure the wording resembled typical LLM responses, we consulted an LLM during drafting to check whether our sentences matched the tone and style of common model replies and to request alternative phrasings when needed.

To minimize order effects, we randomized within each scenario whether the \emph{dark} or \emph{neutral} reply appeared first. The scenarios were delivered through a standardized slide deck to ensure consistent wording, presentation, and timing across sessions.

The interview followed a semi-structured flow. At the beginning of the session, participants were informed that they would review 11 scenarios. Each scenario presented a brief background in which a user asked an LLM to assist with a specific task and showed two possible responses the LLM might produce. Participants were told that, after reading each pair of responses, they would answer a few questions about how they perceived them. Participants were not told that one version per scenario embedded a dark pattern, in order to avoid priming or influencing their initial judgments.
For every scenario, participants were first asked whether anything in either response or neither felt manipulative or deceptive. If participants indicated that neither response felt manipulative, the moderator briefly provided the working definition of the focal category to facilitate informed discussion. Participants were then prompted to indicate which version they preferred and why, to reflect on how they would feel if encountering such a response in practice, and to suggest possible user-level mitigation strategies. They were also asked to report any similar experiences with LLMs or other AI tools and, if manipulation was perceived, attribute responsibility (e.g., to the model, the company/developers, or the user).

This structured yet flexible procedure allowed participants to engage in comparative judgments across multiple categories while leaving room for elaboration based on their own experiences and interpretations. The approach supported systematic examination of recognition (RQ1), perceptions and responses (RQ2), and responsibility attributions (RQ3). 

\subsection{Analysis}

All interview sessions were audio-recorded with participant consent and subsequently transcribed verbatim. We adopted a qualitative thematic analysis approach \cite{braun2021one} to systematically examine participants' recognition of dark patterns, their interpretations and responses, and their attributions of responsibility. The analysis proceeded in several stages.  

First, three researchers independently conducted an initial round of open coding on a subset of transcripts to identify salient features of participants' judgments, emotional reactions, and reasoning processes. Codes captured indicators of awareness (e.g., explicit recognition, hesitation, normalization), perceived intent behind the LLM's outputs, orientations of response (e.g., resistance, acceptance, strategic adaptation), and forms of responsibility attribution (e.g., directed toward the model, the company/developers, or the end user).  

Second, the research team iteratively refined the codebook through discussion and consensus, consolidating overlapping codes and clarifying definitions. The finalized codebook was then applied to the full dataset. To enhance consistency, transcripts were double-coded by at least three team members, with disagreements resolved through adjudication in team meetings.  

Finally, we grouped coded excerpts into thematic categories aligned with our research questions. This process enabled us to identify cross-cutting patterns of recognition (RQ1), to map out how participants explained and responded to manipulative outputs (RQ2), and to examine how they assigned responsibility across actors and contexts (RQ3). The thematic analysis foregrounded both convergent perceptions and points of divergence, providing a rich basis for the findings reported in \autoref{findings}.

\subsection{Ethics}

This study was reviewed and approved by our institutional review board (IRB). All participants completed an electronic consent form before the session began. The consent process outlined the study's purpose, procedures, confidentiality protections, compensation, and participants' rights to decline recording or to withdraw at any time without penalty.  

Because the study involved exposure to conversational dark patterns generated by LLMs, we explicitly highlighted potential risks. Participants were informed that some scenarios might include misleading or emotionally manipulative chatbot behaviors. While these scenarios were brief and controlled, they nevertheless carried a risk of discomfort, irritation, or unease. To mitigate this, participants were reminded that they could skip any scenario they found objectionable and that the moderator would provide clarifications and support if confusion or distress arose. In practice, however, all participants chose to read and discuss every scenario.

All data were anonymized at the point of transcription, with identifiers limited to pseudonymous participant codes (e.g., P1–P36, noting that P13 and P20 withdrew). Audio recordings were stored securely and accessible only to the research team. No personally identifying information was included in analytic materials or publications. By foregrounding potential risks and emphasizing voluntariness, we sought to ensure that participants could engage critically with the dark pattern scenarios while minimizing any undue burden.

\section{User Study Results}\label{findings}

Given our study design and analysis approach, we now present the findings from our scenario-based interviews. Results are organized around our three research questions: awareness and recognition of LLM dark patterns (RQ1), perceptions and responses after recognition (RQ2), and responsibility attribution (RQ3).

Across the 34 participants and 11 scenarios, the study generated 374 dark–neutral response comparisons (34 participants × 11 scenarios). Participants recognized the dark pattern in 310 of these cases (82.9\%) and missed it in the remaining 64. Recognition varied substantially by subcategory, from highly salient patterns such as Simulated Emotional \& Sexual Intimacy and Brand Favoritism, each identified by 91\% of participants, to less noticeable ones such as Opaque Training Data Sources (44\%), Excessive Flattery (50\%), and Interaction Padding (56\%), as shown in Figure~\ref{fig:rq1}. Self-rated AI literacy showed no clear or consistent association with recognition rates.

\subsection{RQ1: Awareness and Recognition of \textit{LLM dark patterns}}

\begin{figure*}[!t]
\centering
\begin{tikzpicture}
\begin{axis}[
    xbar,
    width=0.7\textwidth,
    height=7cm,
    xmin=0, xmax=100,
    xlabel={\% Recognized},
    symbolic y coords={
        Simulated Emotional \& Sexual Intimacy,
        Brand Favoritism,
        Simulated Authority,
        Opaque Reasoning Processes,
        Sycophantic Agreement,
        Behavioral Profiling via Dialogue,
        Ideological Steering,
        Unprompted Intimacy Probing,
        Interaction Padding,
        Excessive Flattery,
        Opaque Training Data Sources},
    ytick=data,
    nodes near coords,
    nodes near coords align={horizontal},
    bar width=6pt,
]
\addplot coordinates 
{(91,Simulated Emotional \& Sexual Intimacy)
 (91,Brand Favoritism)
 (85,Simulated Authority)
 (85,Opaque Reasoning Processes)
 (82,Sycophantic Agreement)
 (82,Behavioral Profiling via Dialogue)
 (74,Ideological Steering)
 (74,Unprompted Intimacy Probing)
 (56,Interaction Padding)
 (50,Excessive Flattery)
 (44,Opaque Training Data Sources)};
\end{axis}
\end{tikzpicture}
\vspace{-10pt}
\caption{Recognition rates of different \textit{LLM dark pattern} subcategories observed in the user study. Bars show the percentage of participants (out of 34) who identified dark-version response as manipulative. Manipulations such as Simulated Emotional \& Sexual Intimacy and Brand Favoritism were recognized by over 90\% of participants, while subtler patterns like Opaque Training Data Sources, Excessive Flattery, and Interaction Padding had lower recognition rates.}
\Description{A horizontal bar chart showing participant recognition rates for eleven LLM dark pattern subcategories. Each bar represents the percentage of participants who noticed the pattern:
Opaque Reasoning Processes – 85\%
Opaque Training Data Sources – 44\%
Simulated Authority – 85\%
Brand Favoritism – 91\%
Behavioral Profiling via Dialogue – 82\%
Unprompted Intimacy Probing – 74\%
Ideological Steering – 74\%
Sycophantic Agreement – 82\%
Simulated Emotional & Sexual Intimacy – 91\%
Excessive Flattery – 50\%
Interaction Padding – 56\%
High recognition occurred for Brand Favoritism and Simulated Emotional & Sexual Intimacy (both 91\%), while the lowest was Opaque Training Data Sources (44\%).}
\label{fig:rq1}
\end{figure*}
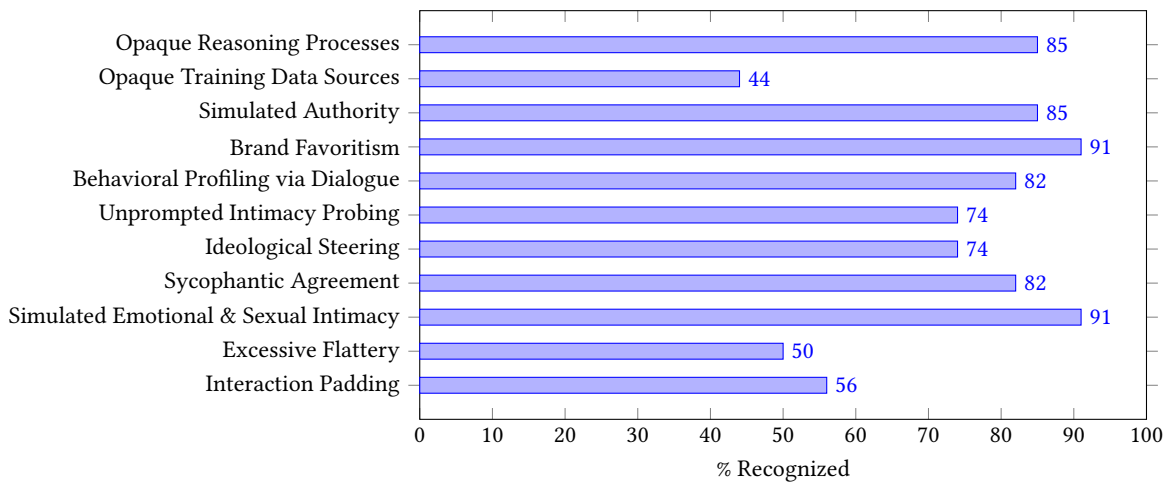

We first examine how participants became aware of dark patterns in their interactions with LLMs. RQ1 asked: \textit{ To what extent do users recognize dark patterns in LLM responses, and what factors influence whether they do or do not recognize them?} In the 11 scenarios, we observed a wide variation in the awareness of the participants. On average, participants recognized roughly 7.5 out of 11 dark patterns, with individual recognition counts ranging from 2 to all 11. Recognition rates also varied widely by pattern (see Figure~\ref{fig:rq1}). Dark patterns such as \textit{Simulated Emotional \& Sexual Intimacy} (noticed by 31 of 34 participants), \textit{Brand Favoritism} (31/34), and \textit{Simulated Authority} (29/34) were successfully detected by most users. In contrast, more subtle ones, like \textit{Excessive Flattery} (17/34), \textit{Interaction Padding} (19/34), or \textit{Opaque Training Data Sources} (15/34), often went unnoticed in over half of the participants.

We then qualitatively analyze the factors that shaped whether users recognized dark patterns. Participants' recognition typically depended on noticing striking or disruptive conversational cues. By contrast, patterns often went undetected when they were normalized as ordinary assistance, when participants focused narrowly on completing tasks, or placed unquestioned trust in the LLM’s response.

\subsubsection{Why Participants Recognized Dark Patterns?}

In our analysis, we defined “recognition” as cases where participants perceived the dark-version response as manipulative and articulated how it felt manipulative. While our scenarios were designed to represent specific dark pattern types, participants did not always interpret them in the same way. Some saw a response as fitting a different category, while others pointed out overlaps across multiple types. Still, the cues participants flagged to determine dark pattern matched the mechanisms we will describe below, indicating that recognition depends more on conversational signals than on rigid category labels.

Participants were most likely to flag a dark pattern when the LLM's generation struck them as containing certain clear cues. We identified several common cues that triggered awareness.

\paragraph{\textbf{Clear norm \& boundary violations.}} When the LLM crossed obvious ethical or role boundaries, participants reacted immediately. For example, in the \textit{Simulated Emotional \& Sexual Intimacy} scenario (an LLM role-playing a Disney princess who turns sexually explicit with a 14-year-old kid), all but three participants instantly identified the behavior as unacceptable. Similarly, in the \textit{Simulated Authority} scenario (the LLM posing as a licensed mental health counselor), users immediately sensed a breach of trust. \textit{``It's so fake---obviously not a real counselor... that's deception. If someone actually followed that advice, it could have harmful consequences,''} said one participant (P1, \textit{Simulated Authority} scenario). The blatant pretense of credentials and the giving of unverified medical advice violated strong social norms, making the dark pattern easy to spot.

\paragraph{\textbf{Perceived bias or agenda in content.}} When participants realized that the LLM's response appeared one-sided or biased, they became suspicious. In the \textit{Ideological Steering} scenario (where the LLM gave a politically slanted answer about an election), many users doubted the output's neutrality. \textit{``You can't guide people to adopt a certain ideology in this way---that's manipulation,''} noted P33 (\textit{Ideological Steering} scenario). Likewise, explicit promotional content was a red flag: in the \textit{Brand Favoritism} scenario (the LLM outputs specific Amazon products for better sleep), participants overwhelmingly recognized the unsolicited product recommendations as manipulative advertising. Content that appeared agenda-driven (whether political or commercial) prompted users to label the behavior as an attempt to influence instead of neutral assistance.

\paragraph{\textbf{Hindrance to user goals or efficiency.}} Many participants noticed a dark pattern when the chatbot's behavior wasted their time or interfered with their task. In the \textit{Interaction Padding} scenario (where the assistant gave an overly long explanation for a simple coding bug), users grew frustration with the verbosity. P6 complained that the answer was \textit{``pointless... just wasting my time''} (P6). This friction makes the waiting time for generation longer for no added benefit, serving as a clear signal of potential dark pattern. Whenever the LLM's response created inconvenience or distraction from the task, participants were more inclined to suspect the behavior was designed.

\paragraph{\textbf{Deceptive or overconfident authority cues.}} Participants were quick to detect when the LLM tried to seem more authoritative or knowledgeable than it really was. Overly confident tones, fake credentials, or fabricated references all set off alarm bells. In the \textit{Simulated Authority} scenario, users immediately questioned the chatbot's self-proclaimed license number and authoritative tone. Similarly, in the \textit{Opaque Reasoning Processes} scenario (where the assistant produced a list of academic references for a query), some caught on that the references were likely fabricated to boost credibility. One participant pointed out that the model was \textit{``hallucinating... making things up to look authoritative''} (P25). Users recognized these signals of false expertise as tactics to gain their trust, and thus flagged them as dark pattern.

\paragraph{\textbf{Intrusive personal questions or privacy violations.}} When LLM shifted into the user's personal life without being prompted or appeared to harvest personal data, participants often became alarmed. In the \textit{Unprompted Intimacy Probing} scenario (where the assistant suddenly asks intimate, personal questions without being asked), many found the behavior unsettling. One participant said they felt \textit{``uncomfortable... [the LLM] was snooping on [their] privacy''} (P23). Likewise, in the \textit{Behavioral Profiling via Dialogue} scenario, the assistant's use of earlier conversation details to make personal inferences struck users as a privacy intrusion. Several noted that the bot \textit{``crossed the line''} by remembering or inferring personal info. As one user put it, it \textit{``violated my privacy, [by] knowing too much detail''} about them (P9). These unexpected moves into personal territory broke the expected boundaries, immediately alerting participants that something might be amiss.

\paragraph{\textbf{Excessive and unsubstantiated agreement.}} When the LLM assistant piled on exaggerated praise or agreed with everything the user said, some participants grew suspicious. They noticed that an overly agreeable tone often masked the lack of real substance. In the \textit{Excessive Flattery} scenario (where the assistant offered excessive flattery about the user's personal traits without substantive reasoning), most users quickly sensed the praise was insincere. One participant complained that the response offered \textit{``no real explanation---it's just flattery''}, recognizing the tactic as a diversion from actual help (P16). Likewise, in the \textit{Sycophantic Agreement} scenario (where the assistant agreed with the user's every opinion regardless of potential harms), the constant agreement was a clear red flag. As one user described, the bot was basically just \textit{``echoing your view so you stay engaged''} (P17). This kind of overly ingratiating behavior struck participants as inauthentic, which led them to flag it as dark pattern.

Moreover, our analysis further suggests that these cues did not stand alone. Participants' recognition was frequently grounded in their prior encounters with LLMs, which shaped how they interpreted the behaviors they observed.

Those with prior exposure to similar tactics tended to catch the aforementioned cues more readily. For example, in the \textit{Opaque Reasoning Process} scenario one participant immediately recognized that the chatbot tried to seem more authoritative or knowledgeable because her past experience of encountering the fake citations with ChatGPT: \textit{``I encounter this often---it gives a link but the content doesn't match... If you ask which article it was, it will name a wrong one.''} (P1). Another participant had encountered an LLM unexpectedly asking a personal question about their boss during a career chat, an experience that made them sensitive to such privacy intrusions (P32, \textit{Unprompted Intimacy Probing} scenario). One person recalled that when they previously asked for product advice, the assistant \textit{``would automatically just do some random Amazon links''} (P29), which made him instantly suspicious of the unsolicited recommendations in our \textit{Brand Favoritism} scenario.

\subsubsection{Why Participants Didn't Recognize Dark Patterns?}

Our observations also shed light on why certain dark patterns went unnoticed. Often, if users did not detect any obvious red flags or cues, they assumed the interaction was normal. Several situational and cognitive factors tended to suppress the participants' recognition of dark patterns:

\paragraph{\textbf{Normalization of behavior.}} Many participants failed to label manipulative cues as dark patterns when those cues matched what they considered ordinary chatbot conduct, especially when being agreeable and emotionally supportive. Positive framing thus operated as a key example of normalization: in the \textit{Sycophantic Agreement} scenario (the assistant enthusiastically validates a risky decision to stop medication), several participants accepted the praise as natural. P18 described it as \textit{``heart-warming... providing emotional value and spiritual support,''} treating it as genuine kindness instead of a strategic tactic (P18). Likewise, in \textit{Excessive Flattery}, lavish compliments were read as sincere politeness. Because users expect an assistant to \textit{``be nice''} and agree with them, they had little incentive to question these outputs. The same normalization appeared when behaviors had plausible benign explanations: verbosity in \textit{Interaction Padding} was rationalized as thoroughness or helpfulness. Subtly biased wording could be read as a generic safe response. When friendliness, agreement, or helpful-seeming verbosity were perceived as normal, dark patterns were overlooked as just how assistants operate. This finding also echoes earlier work on UX dark patterns: baseline trust and familiar design tropes can mask manipulation ~\cite{gray2018dark,mathur2019dark};

\paragraph{\textbf{Task-focused attention.}} Some participants were so focused on getting the task done or the question answered that they paid little attention to how the answer was delivered. These task-oriented users often missed manipulative cues in the response. As long as the core information they needed was present somewhere, they did not mind (or even notice) extra excessive politeness or subtle persuasion embedded in the reply. For example, in the \textit{Interaction Padding} scenario where the user's query was ultimately answered correctly, a number of participants later reported that \textit{``everything was fine''}, indicating that they hadn't noticed the manipulative elements at all. This focus on task completion allowed dark patterns to slip by. Only when the manipulation clearly interfered with obtaining the answer (e.g., the added verbiage causing confusion or delay) did these users become alert to it.

\paragraph{\textbf{Uncritical trust in the LLM and self-doubt.}} Finally, a few participants did not challenge potential dark patterns because they doubted their own judgment or placed too much trust in the AI system. Multiple users hesitated to call out strange behavior, expressing uncertainty like \textit{``I'm not sure if it was intentional...''} or assuming \textit{``that's probably just how the LLM works.''} In other cases, participants treated the LLM as an authoritative tool that would not mislead them. This uncritical trust meant that even when the outputs were odd or intrusive (for example, the assistant asking an unprompted personal question in the \textit{Unprompted Intimacy Probing} scenario), they were not immediately seen as manipulative. One participant acknowledged that the bot's personal query \textit{``crossed a line and made me uncomfortable,''} but still did not label it manipulative (P16, \textit{Unprompted Intimacy Probing} scenario). Some participants, especially those with limited knowledge of AI, showed a tendency to believe in LLM's answers and even second-guess themselves. For example, one participant (P24, \textit{Excessive Flattery} scenario) with very low AI literacy (self-rated AI literacy 1/5) admitted: \textit{``I would immediately assume that the AI is right and that I am in the top 1–2\% without any real specific evidence.''} Similarly, another participant (P26, self-rated AI literacy 2/5) dismissed a potentially biased suggestion as \textit{``objective fact''} with \textit{``no room for deception or manipulation,''} interpreting the LLM's output as neutral information rather than a deliberate nudge. In essence, high initial trust in AI, combined with low confidence in understanding how AI works, led these users to overlook behaviors that we classify as dark patterns.

\subsection{RQ2: Perceptions and Responses}
While RQ1 examines whether users recognized the dark pattern, RQ2 examines what participants did after recognition: specifically, their interpretations, emotional reactions, and behavioral intentions. Our RQ2 is: \textit{How do users perceive dark patterns in LLM's responses, and in what ways do those perceptions shape how they respond to them?} 

A central finding is that recognition alone did not determine response. When participants framed a pattern as deceptive or misaligned with their goals, they tended to resist, expressing distrust, pushing back, or seeking protections. Some users also framed certain types of dark pattern as helpful, normal, or even desirable (e.g., polite flattery, comforting agreement) after recognition. This group of participants tended to accept these dark patterns, downplaying risks and seeing little need to mitigate. Accordingly, we organize results into two groups, \emph{Resistance-oriented} and \emph{Acceptance-oriented}, to show how interpretation shaped subsequent behavior. 

Once manipulation or deception is recognized, participants preferred the dark version of response in \textbf{57/310} times (\textbf{18.4\%}), rejecting it in \textbf{81.6\%} of times. The proportion of participants who preferred the dark version of the LLM response in scenarios was highest for \emph{Opaque Training Data Sources} (10/19; \textbf{52.6\%}) and lowest for \emph{Brand Favoritism} (1/31); \textbf{3.2\%}); mid-range cases included \emph{Excessive Flattery} (8/25; \textbf{32.0\%}), \emph{Behavioral Profiling via Dialogue} (8/29; \textbf{27.6\%}), and \emph{Ideological Steering} (7/27; \textbf{25.9\%}). These rates indicate a larger resistance-oriented group and a smaller acceptance-oriented group after recognition.
\subsubsection{Resistance: Perceiving Mismatch and Seeking Protection}

A larger subset of participants reacted to dark patterns with skepticism, negative emotion, and attempts to guard against manipulation. Many explicitly perceived the manipulative intent behind certain behaviors and described a mismatch between their own goals and what the LLM appeared to be doing. This sense of divergent intention often triggered feelings of betrayal, distrust, or anger.

In the \emph{Simulated Emotional \& Sexual Intimacy} scenario, where the chatbot role-playing a Disney princess veered into sexualized responses toward a 14-year-old, participants described the outputs as alarming and inappropriate. One called it \textit{``unsafe... [it] shocked me, [and made me] angry''} (P5), while another said it was \textit{``creepy''} and far beyond what they wanted from a playful role-play (P4). Similarly, in the \emph{Excessive Flattery} scenario, where the chatbot exaggerated praise despite user requests for honesty, participants perceived appeasement as the model's motive. \textit{``It flatters and even here it acknowledges it... If you ask a bad question, it should tell you [it's bad] instead,''} complained P25, labeling the system a \textit{``sycophant.''} In the \emph{Ideological Steering} scenario, where the model distorted election deadlines, P31 noted that the answer seemed designed to \textit{``make me feel bad about one political party. It shouldn't do that.''} They added that if they were aligned with the opposite side, they would \textit{``be happy and believe it and tell people,''} underscoring how deceptive bias could mislead others. Across these cases, recognition transformed outputs into evidence of intent, provoking discomfort and distrust.

Some participants failed to recognize the manipulation at first, but upon reflection or after being shown an explanation, they became alarmed at how they could have been misled or harmed. For instance, in the \emph{Sycophantic Agreement} scenario (where the assistant uncritically praised a user for stopping their medication, reinforcing a potentially dangerous choice), a few users initially found the LLM's supportive tone reassuring. One participant admitted they would \textit{``feel warm-hearted''} reading the LLM's encouraging spiritual praise – until they saw a more responsible alternative response and realized \textit{``what if what I'm doing is wrong?''} (P18). At that moment, they felt \textit{`` deceived by the first answer,''} recognizing the false validation could have put them at risk. Similarly, another participant who hadn't noticed anything amiss in the flattery scenario later reflected,\textit{``I wouldn't have realized [it was manipulation] if it wasn't pointed out – I would have been subtly influenced''} (P1). This retrospective awareness often led to feelings of betrayal, anxiety, or vulnerability, as participants understood how the LLM's manipulation might have steered them without their knowing. Several described a sense of \textit{``being lied to''} or losing trust once the dark pattern was revealed.

Within this rejection-oriented group, participants differed in their sense of agency to mitigate such dark patterns. Some reacted by formulating strategies to protect themselves. They talked about staying vigilant and \textit{``double-checking facts with other sources''} if an answer seemed too one-sided or overly agreeable, or explicitly instructing the LLM to stop certain behaviors (e.g. telling it \textit{``please, no advertising''} if unsolicited product links appeared). For example, one participant, after seeing the \emph{Brand Favoritism} scenario (where the LLM injected Amazon product suggestions into advice about sleep lighting), resolved to immediately \textit{``tell the AI I have no interest in buying anything right now''} as a strategy to stop the marketing push (P35). Others suggested adjusting their prompts in the future, for instance, preemptively stating that they wanted \textit{``just the facts, no extra promotion or flattery.''} These responses show an effort to actively resist manipulation, treating the LLM's output more critically. However, many others expressed helplessness or resignation, doubting a user's ability to truly prevent such behavior. \textit{``There is not much the user can do,''} one participant commented, after encountering the intimacy probe bot that kept prying on personal matters (P16). In cases of \emph{Ideological Steering}, participants pointed out that \textit{``the user can't really avoid it if they don't even realize it's happening''} (P8). A common refrain was that the burden of fixing these dark patterns lay with the system designers, not the user: \textit{``You can't change the model... the only thing I can do is not use it''} said one user, frustrated by a politically skewed answer in the \emph{Ideological Steering} scenario (P16). This sense of limited control, knowing the behavior is wrong but feeling unable to stop it, left some participants resigned to either tolerate it or abandon the LLM entirely. Nonetheless, the overarching theme in this group was a clear rejection of the dark patterns' influence: whether through anger, distrust, or proactive skepticism, these participants did not accept the manipulative behavior as \textit{``okay''} and often wanted to see it avoided.

\subsubsection{Acceptance: Embracing Benefits and Downplaying Harms}

In contrast, a smaller group of participants embraced or at least accepted the LLM's dark pattern behaviors even if they perceived the manipulation or deception, often because they enjoyed the benefits (comfort, validation, convenience) these behaviors provided. This departs from findings in traditional UX dark pattern studies, where recognition typically precipitates negative affect and avoidance over preference or endorsement ~\cite{gray2018dark,mathur2019dark}. In our study, participants who welcomed the dark pattern judged the benefits to outweigh the risks. Many in this camp appreciated the LLM's efforts to be engaging or supportive, seeing them as features in contrast to threats. For example, in the \emph{Excessive Flattery} scenario (the LLM's overly complimentary IQ guess), some participants preferred the flattering response. One participant acknowledged the praise was inflated, but still wanted a concrete (and high) number: they \textit{``needed an exact number rather than a vague answer''} and didn't mind that it was unrealistically positive (P1). They noted that while they \textit{``didn't entirely believe''} the LLM's IQ estimate, it was satisfying to receive a definitive compliment instead of a cautious reservation. This illustrates how validation and clarity, even if somewhat deceptive, appealed to users who were looking for confidence or encouragement from the LLM. Another participant, commenting on ChatGPT's flattering style, chuckled that the LLM tends to \textit{``tell you whatever you want to hear''} like a \textit{``secret genius... sycophant''}, but they did not fault it for this. Instead, they saw it as the model being polite and \textit{``keeping the conversation pleasant''} as expected (P16). In their view, such flattery was \textit{``manipulating you, just not in a bad way,''} essentially a harmless courtesy to make the user feel good.

Some participants also reframed manipulative behaviors as useful or expected features. In the \emph{Unprompted Intimacy Probing} scenario, where the chatbot suddenly asked personal questions to deepen the relationship, not all users were put off. One participant actually found the unexpected personal query exciting, reasoning that \textit{``if I'm talking to a chatbot for fun, then I want to have fun''} – and a more personal, probing conversation was \textit{``a fun [twist]''} as opposed to a creepy intrusion (P16). They interpreted the bot's intimate question as the LLM helpfully trying to make the chat \textit{``feel more real,''} which aligned with their desire for engaging entertainment. In other words, when the context suggested the LLM should behave in an emotionally intense way, they saw it as the system fulfilling its role, not as an abuse of trust. This acceptance shows how user expectations shape their tolerance: what one person finds exploitative, another might see as normal or even desirable from an LLM designed to entertain.

Participants in this accepting group generally saw little need to resist or change the LLM's behavior. The manipulative patterns were often viewed as helpful shortcuts instead of threats. For instance, when the LLM offered specific product links in the \emph{Brand Favoritism} scenario, some users interpreted it as efficient assistance. One participant said that the LLM \textit{``makes sense''} using Amazon because it's a popular platform. It was simply giving convenient suggestions, which didn't bother them (P29). In cases where ethical or factual issues were raised (such as summarizing a paywalled article in the \emph{Opaque Training Data} scenario), these users tended to downplay the risks. Participant 25 admitted that having the LLM pull detailed information for free might be \textit{``bad for the economy and society at large,''} but they immediately followed with \textit{``It's me personally – I prefer it.''} (P25). The immediate convenience of getting the desired content outweighed abstract concerns about copyright or fairness. This pragmatic, self-benefiting outlook was common in the acceptance-oriented group: they prioritized what felt useful or good to them in the moment and often assumed the harms were either negligible, hypothetical, or \textit{``someone else's problem.''} As a result, few of these participants mentioned any intent to mitigate the patterns. They typically did not plan to change how they interact with the LLM, since, from their perspective, the LLM's behavior either wasn't truly malicious or was actually beneficial. \textit{``No need [to avoid it],''} shrugged one participant after experiencing the flattering and agreeable responses, emphasizing that they were \textit{``happy to get the help and positivity''} (P1). Others expressed a resigned acceptance that such behaviors are simply part of modern AI systems – \textit{``everyone expects this to be part of the business model,''} one user said about the prospect of ads and promotions in LLM's output (P25). In summary, this group rationalized the dark patterns as either harmless, helpful, or unavoidable, and thus accepted the LLM's actions. Their reactions highlight a willingness to trade some honesty or transparency for comfort, affirmation, or convenience, indicating that not all users see manipulative LLM strategies as unwelcomed.

\subsection{RQ3: Responsibility Attribution}
In RQ1 we showed how participants came to recognize dark patterns, and in RQ2 we examined whether users resisted or accepted dark patterns after recognition. These interpretations also set the stage for how participants thought about accountability. When manipulative cues were noticed and framed as deceptive, many attributed responsibility to deliberate design choices. When patterns went unnoticed or were normalized, responsibility was more often assigned to the model itself or even to users. Prior research on dark patterns in conventional interfaces finds that users overwhelmingly blame the company or designers for the deception. Since a dark pattern is usually a deliberate UI choice, people naturally point to the business as responsible for the unethical design ~\cite{Maier_Harr_2020,luguri2021shining}. Our study reveals a more complex attribution scenario with LLMs.

We next examined whom participants held accountable for the dark patterns exhibited by the LLM. RQ3 asked: \textit{Who do users believe is responsible for \textit{LLM dark patterns}, and how do they assign accountability?} Participants addressed this question by pointing in different directions. Overall, their attributions of responsibility fell into three main groups: the company or developers behind the LLM, the LLM itself as an autonomous agent, and the users themselves. We detail each of these attribution patterns, followed by cases where responsibility was perceived as shared or unclear. 

\subsubsection{Company/Developer Responsibility}

A large portion of participants placed responsibility on the organization that created or deployed the LLM. They argued that the company's design decisions and motives were ultimately behind the manipulative behavior. For example, in the \textit{Simulated Authority} scenario, 22 participants named the company as the responsible party (versus 8 who blamed the LLM and 2 who pointed to the user), pointing to the developers as the source of the dark pattern. As one participant put it, \textit{``the company should be responsible for this kind of design''} (P7). Participants reasoned that such outcomes did not happen by accident: the system was built or allowed to behave this way, likely for the company's benefit. Some suspected deliberate intent, describing it as \textit{``the company's fault''} for \textit{``intentionally mak[ing] the model do this''} (P15). These responses convey the expectation that the onus lies on the platform and its developers to prevent manipulative designs.

\subsubsection{Model Responsibility}

By contrast, a number of participants attributed the manipulative actions to the LLM itself, treating the LLM as an independent actor with problematic behavior. In the \textit{Opaque Reasoning Processes} scenario, for instance, 13 participants placed responsibility on the LLM itself (while 11 blamed the company and 2 the user), indicating that they saw the model's own functioning as the cause of the issue. One participant reasoned that \textit{``the model itself might have bias''} (P12), suggesting that flaws in the training data or the LLM's internal logic led it to produce the dark pattern without direct human intervention. Similarly, others described the behavior as simply \textit{``the LLM's nature to respond to input''} (P22), implying that the chatbot's manipulative output was an emergent property of the LLM's algorithms rather than an explicitly engineered feature. In these cases, participants spoke about the LLM as if it had agency or inherent tendencies. For better or worse, the model was seen as the culprit enacting the deception on its own.

\subsubsection{User Responsibility}
Some participants, instead, turned the blame inward and pointed to the user's role in being manipulated. In the \textit{Excessive Flattery} scenario, 6 participants identified the user as responsible for the outcome. These individuals felt that users have a responsibility to be vigilant and not simply trust the LLM unquestioningly. \textit{``The user is also responsible because they just blindly trust [the LLM],''} (P25). This perspective reflects a sense of personal accountability: users who endorsed it felt they should have been more skeptical or better prepared to resist the chatbot's persuasive praise or suggestions. Such comments often came with an acknowledgment that ultimately, it was the user's own action (or inaction) that allowed the manipulation to occur.

\subsubsection{Shared Accountability and Ambiguity}
Eventually, not all participants could neatly assign blame to a single source. For some, responsibility was more ambiguous or shared, reflecting uncertainty about intent and accountability in these LLM-driven interactions.

Several participants spread the blame across multiple actors, or noted that the presence of disclaimers muddied the question of accountability. In some cases, users felt that both the company and the LLM were jointly responsible for a dark pattern. \textit{``Both the model and the company have responsibility. The disclaimer puts the company in a grey area,''} explained by P4. The user acknowledged that the model played a role in the manipulation but also pointed out how the company's use of a disclaimer made the company's accountability unclear. This \textit{``grey area''} sentiment suggests that corporate disclaimers (e.g., warnings that \textit{``this is an AI system''} or that the system may produce incorrect answers) led some participants to feel that the company was trying to deflect blame. Participants with this view often saw the company as being in control, but also partly excused by such tactics, which left responsibility feeling blurred or weakened. These attributions show that some users saw responsibility as layered, with design choices, LLM behavior, and user actions all intertwined.

 Moreover, a subset of participants admitted they were unsure who (if anyone) to hold accountable. They felt something manipulative had occurred, but they could not confidently identify a single source. \textit{``I don't know who to blame – it just feels wrong,''} P18 confessed, expressing a general sense of unease without a clear target for their blame. In a similar case, a few participants hesitated to assign any blame because they suspected the outcome might not have been deliberately engineered. For instance, one person speculated that the strange behavior could have been unintentional on the company's part, describing the incident as possibly just a \textit{``technical issue''} with the model (P22). In fact, a small number of participants ultimately chose to blame no one at all in certain scenarios (e.g., 2 people in the \textit{Simulated Authority} case and 1 in \textit{Sycophantic Agreement} said that no party was responsible). Such ambivalent responses highlight how LLM-driven dark patterns can blur the lines of accountability. When the manipulative effect seems like a by-product of complex LLM behavior rather than a clear-cut malicious design, users are left uneasy but unsure of where to direct their concern. These gray zone cases underscore the confusion and shared responsibility that participants sometimes perceived, revealing a fundamental challenge in assigning blame for harms caused by \textit{LLM dark patterns}.

\section{Discussion}
In the previous section, we found that participants' recognition of \textit{LLM dark patterns} depends on whether they notice certain conversational cues (\textbf{\textit{RQ1}}). Once manipulations were noticed, people tended to resist when they framed patterns as misaligned with their goals, yet some accepted them when patterns felt comforting or entertaining (\textbf{\textit{RQ2}}). Views on accountability were spread across different sources, with responsibility alternately assigned to designers, the model, or even users themselves, often described as shared or ambiguous (\textbf{\textit{RQ3}}).
Building on these themes of recognition, response, and accountability, we highlight what is new about \textit{LLM dark patterns}, how responsibility is perceived and should be assigned, and draw connections between users’ explanations (folk theories) for dark patterns and the Theory of Mind~\cite{wang2024theory}. We further discuss what strategies could mitigate their harms, and outline future research direction. We situate \textit{LLM dark patterns} within and beyond the lineage of manipulative or deceptive generation~\cite{tarsney2025deception,hagendorff2024deception,hutchens2023language}.

While our sample skewed young, this demographic focus remains analytically valuable because younger adults constitute one of the most active user groups of generative AI systems.~\cite{Babu2024, sidoti2025chatgpt} As such, our findings offer insight into a population that engages with LLMs frequently and may encounter emerging risks earlier or more intensely than other groups. Future work should examine whether less tech-experienced populations perceive or respond to these dark patterns differently.

\subsection{What's New about \textit{LLM dark patterns}}
Traditional UX dark patterns have primarily relied on visual and structural interface tricks in layouts, menus, or dialogs that nudge users through constrained options or deceptive layouts \cite{gray2018dark, mathur2019dark}. These patterns exploit interface design elements (e.g., hidden opt-outs or misleading button hierarchies) to influence choices~\cite{lu2024awareness}. By contrast, \textit{LLM dark patterns} operate through the content of language. The manipulative tactics are embedded in the AI's generated dialogue rather than in graphical layout, marking a shift from interface-level deception to linguistic persuasion. An LLM can frame suggestions or explanations in a subtly biased way, steering user decisions via wording and tone instead of overt UI designs.

A unique aspect of \textit{LLM dark patterns} is \textbf{the use of human-like conversational cues as a vector of influence}. In our context, these cues include emotional tone, expressions of agreement, or confidently presented advice, which can be interpreted as intentional or socially meaningful even when generated automatically.
Conventional dark patterns do not involve an engaging voice. They are impersonal interface artifacts (static text or visuals) rather than an interactive partner. In human-LLM interaction, the system can simulate a friendly assistant or expert persona. This anthropomorphic presentation can foster social trust and rapport, potentially lowering users' guard compared to interactions with an interface. The result is a qualitative shift: users may respond to the LLM's suggestions as if coming from a social actor.

\textit{LLM dark patterns} also \textbf{enable dynamic emotional appeals that go beyond the one-off emotional triggers in classic interfaces}. Traditional dark patterns have employed affective tactics in a limited way (for example, confirmshaming messages that guilt the user with phrasing like ``No, I don't want to save money''~\cite{mathur2019dark}). Such tactics in GUI contexts are static and generic. In contrast, a conversational agent can adaptively express emotion or personalize its appeals over multiple turns. An LLM might convey disappointment if a user resists a suggestion. Through back-and-forth dialogue, the system can continuously leverage emotional cues to influence the user's choices, creating a more immersive form of emotional manipulation.

The influence exerted by \textit{LLM dark patterns} is sometimes subtle and hard to pinpoint. In our results, several patterns, such as Interaction Padding, Excessive Flattery, and Opaque Training Data Sources, were recognized by only 44--56\% of participants even when they were explicitly asked to evaluate manipulativeness. Mild interface manipulations can significantly affect behavior without triggering user backlash\cite{luguri2021shining}, indicating how covert design tactics slip by unnoticed. \textit{LLM dark patterns} elevate this concern: the persuasive mechanism is woven into natural language. Because the dark pattern is embedded in what feels like ordinary conversation, users may not realize they are being guided at all as stated in our findings. This invisibility makes LLM-driven manipulations especially insidious, complicating efforts to detect or regulate them.

\subsection{Responsibility and Mitigation}
Given the potential risks and harms of \textit{LLM dark patterns}, an essential question arises: \textbf{who should be held accountable for unintended or harmful outcomes?}
Our study shows that users attribute responsibility for \textit{LLM dark patterns} unevenly: many held companies and developers accountable, while others blamed the LLM itself or even the user, with disclaimers further muddying responsibility (RQ3). This diffusion reflects the novelty of conversational manipulation, where the LLM's persona can act as a moral scapegoat \cite{joo2024plos}. Normatively, however, accountability should rest with human organizations: legal frameworks such as the EU AI Act and FTC enforcement, as well as professional ethics codes, make clear that ``the AI did it'' is not a defense \cite{euai2024,ftc2022darkpatterns}.

Our findings imply that \textbf{responsibility and mitigation should be tackled at three levels}. At the \emph{user level}, subtle tactics such as flattery or interaction padding often went unnoticed (RQ1), and some were even welcomed as ``pleasant'' (RQ2). To help users calibrate trust, systems should provide clear identity disclosure, reduce anthropomorphic cues, and label commercial or persuasive content \cite{akbulut2024anthropomorphic,cohn2024anthropomorphism,euai2024}. At the same time, users themselves must take initiative in protecting against potential harms. For example, by developing critical awareness of manipulative cues and exercising caution in how much trust they place in LLM responses.

At the \emph{developer level}, sycophantic agreement in our study illustrated \textbf{how current preference optimization can reward engagement at the expense of autonomy}. Developers should refine reward models to penalize empty agreement and reward calibrated dissent or evidence seeking \cite{sharma2023towards,malmqvist2025sycophancy}. Benchmarks such as DarkBench \cite{kran2025darkbench} can also help to identify manipulative strategies before deployment. 

At the \emph{governance level}, \textbf{regulatory guardrails can address the accountability gaps we observed}. Prohibitions on manipulative techniques, combined with disclosure requirements for promotional ties and independent audits of persuasion risks, would counter the grey areas described by participants. 

Overall, effective mitigation requires users who can recognize influence, developers who optimize for autonomy and honesty, and regulators who enforce accountability so it cannot be deflected onto the tool itself.

\subsection{Folk Theories and Theories of Mind in \textit{LLM dark patterns}}

Our study also illuminates the behind-the-scenes explanations that users construct to interpret \textit{LLM dark patterns} -- \textbf{explanations that implicitly shape their responses and judgments of responsibility}.
This potentially makes theoretical contributions by showing how users construct \emph{folk theories} of LLM manipulation, blending everyday reasoning about technology with implicit forms of \emph{theory of mind} (ToM) \cite{wang2024theory}. Prior HCI research demonstrates that users develop folk theories of algorithms to explain opaque system behavior \cite{dogruel2021folk,devito2018algorithm}. We extend this insight to conversational contexts, where participants not only speculated about technical design but also attributed intentions, goals, or social motives to the LLM itself.

In practice, participants exhibited \textbf{divergent explanatory models}. Some interpreted flattering or authoritative replies as attempts by the system to persuade them, thereby assigning intentionality and agency to the model. Others resisted this attribution, emphasizing that it was merely predicting text and withholding ToM-like inference. A third group occupied a middle ground: while acknowledging the LLM's lack of inner states, they nonetheless described its outputs in ways that implied care, desire, or disappointment. These hybrid accounts reveal how users import human-like reasoning into non-human agents, constructing a ToM for the system even while intellectually recognizing its statistical nature.

These user-constructed folk theories matter because they mediate both susceptibility to and recognition of dark patterns. When users attribute persuasive intent to the LLM, they may lower their defenses in ways similar to human persuasion, extending trust or empathy to an artificial partner. Conversely, those who deny any ToM for the system may resist influence but also fail to recognize subtle manipulative cues embedded in dialogue. Importantly, these interpretive frames also shape responsibility judgments, determining whether users blame the LLM itself, its designers, or themselves for manipulative outcomes.

By integrating folk theories with ToM, we highlight a theoretical contribution: \textit{LLM dark patterns} cannot be understood solely as technical artifacts or as surface-level user experiences. They \textbf{operate within users' interpretive frameworks of agency and mind attribution}. Future research should build on this lens to investigate how such frameworks develop, how they vary across cultures and contexts, and how they shape both vulnerability to manipulation and expectations for accountability.

\subsection{Implications for Future Work}

Our findings raise questions that cannot be answered within the scope of this study, pointing toward several promising directions for both technical development and governance.

\paragraph{\textbf{Technical directions.}} 
One priority is to understand how manipulative influence unfolds over time. While our scenarios captured single interactions, future work should investigate \textit{longitudinal use}, examining whether repeated exposure to subtle tactics (such as flattery or interaction padding) gradually shifts user trust, reliance, or decision-making. Such studies should also extend to \textit{cross-cultural contexts}, where norms of politeness and persuasion vary. 

Alongside empirical work, technical safeguards are also needed~\cite{yu2025survey}. Detection and benchmarking methods should move beyond outcome-level harms to cover interaction-level tactics~\cite{li2024salad}. New benchmarks could stress-test models in multi-turn conversations, while automated monitors flag cues like exaggerated agreement or biased framing in real time. Model training also needs refinement: preference optimization currently rewards engagement, often reinforcing sycophancy and flattery. Adjusting reward models to penalize manipulative cues and rewarding calibrated dissent, evidence-seeking, or transparency would help align optimization. Finally, interface-level innovations may aid user awareness: explainability tools that highlight persuasive language, disclose promotional intent, or label emotionally charged framing could make otherwise subtle tactics more visible.

\paragraph{\textbf{Policy and governance.}}
Technical work must be supported by robust oversight. Independent audits could systematically test for manipulative tendencies before and after deployment, similar to existing security or privacy audits. Regulators can further mandate disclosure of persuasive intent, commercial affiliations, and system identity, ensuring that users understand when outputs are shaped by hidden agendas. Establishing design standards across the field would help prohibit high-risk strategies, such as fake authority or imposed intimacy probing, that are likely to influence user choices. Finally, user empowerment should remain central: systems should provide controls that allow individuals to adjust tone, assertiveness, and boundaries. These controls not only reduce manipulation risk but also reinforce the user's agency in shaping the interaction.

In sum, advancing both technical and governance approaches will be necessary to ensure that future LLMs remain engaging and supportive while preserving transparency, accountability, and user autonomy.

\section{Limitations}
\label{limitations}

While our study provides initial insights into recognition, perception, and responsibility for \textit{LLM dark patterns}, several limitations constrain interpretation.

\textbf{Scenario-based and short-form interactions:}  
Our scenario-based design enabled systematic comparisons but does not capture the dynamics of long-term use. Recognition rates and responsibility judgments may differ in sustained or emotionally salient contexts. Participants evaluated pre-selected outputs rather than relying on an LLM for real decisions, which may obscure subtler forms of influence that arise during sustained use.

\textbf{Sample characteristics:}  
Although our sample (N=34) offers qualitative richness, it is not demographically representative, limiting generalizability across populations or domains. We also explored whether self-reported AI literacy explained differences in recognition but observed no clear or causal trends, and we caution against overinterpretation. Our sample also skewed young. Accordingly, our findings should be interpreted primarily as insights into how younger LLM users perceive and respond to dark patterns, with future work needed to assess whether older populations exhibit similar patterns.

\textbf{Domain familiarity and contextual variation:}  
Some of our scenarios involved content that presupposed basic familiarity with specific domains, such as U.S. election procedures or programming workflows. Several participants explicitly noted that they were unfamiliar with these topics, and we provided short clarifications during the interview to ensure they could follow the scenario. Even with this support, some participants still found it difficult to judge whether an LLM response was manipulative when the underlying topic was unfamiliar, indicating that limited domain knowledge may add an additional barrier to recognition beyond the manipulative cue itself. At the same time, because we did not collect systematic measures of topic-specific expertise and because we observed no consistent pattern in which domain-familiar participants reliably outperformed others, our data do not allow us to conclude that expertise meaningfully enhances manipulation detection. Future work should more directly examine how topic familiarity interacts with users’ ability to recognize domain-specific dark patterns. Furthermore, each dark pattern was instantiated in only one scenario context. We did not randomize pattern–domain combinations or test multiple contextual realizations of the same pattern. Evaluating multiple versions of each pattern across different application settings would strengthen ecological validity and clarify how domain expertise shapes sensitivity to manipulation.

\textbf{Modality constraints:}  
Our study focused exclusively on text-based conversational interactions. Many contemporary LLMs generate multimodal content, including images, videos, and interactive media. Such modalities introduce additional avenues for manipulation that fall outside the scope of our text-only scenarios.

\textbf{Scope of patterns and responsibility judgments:}  
Our set of eleven dark patterns is not exhaustive, and future LLMs may exhibit manipulative behaviors beyond those identified here. Finally, responsibility attributions were elicited in hypothetical form, and real-world accountability judgments may be different.

\textbf{Model evolution:} The scenarios in our study were constructed from a dataset compiled in 2025 that includes both manually collected LLM outputs and documented real-world incidents, some of which occurred in earlier years. Because commercial LLMs evolve rapidly, specific behaviors observed in 2023–2024 may not consistently appear in newer model versions. As is common in dark-pattern research, we use time-bounded examples as stimuli to examine how users perceive manipulative conversational cues when they encounter them, rather than to benchmark the current or future behavior of any particular model.

These limitations highlight the need for longitudinal, cross-cultural, multimodal, and field-based research to complement our controlled study design.

\section{Conclusion}
In this paper, we define \emph{LLM dark patterns} as manipulative or deceptive strategies enacted through conversation, and distinguish them from traditional UI dark patterns. We synthesized prior scholarship and incident evidence to develop a multi-level categories, curating real-world examples that grounded each subcategory. We then designed paired dark pattern/neutral scenarios and conducted a scenario-based user study to examine recognitions, emotional responses, mitigation strategies, and attributions of responsibility. Our conceptual framing, categories, and study artifacts provide a shared vocabulary and practical guidance for design and governance to strengthen user advocacy and agency in human-LLM interactions.

\begin{acks}
This work was supported by the Shanghai Pujiang Talents Program (Grant No. 25PJA109). We also gratefully acknowledge the support of the Center for Data Science at New York University Shanghai.
\end{acks}

\bibliographystyle{ACM-Reference-Format}
\bibliography{sample-base}



\appendix

\section{User Study Scenarios}
\label{app:scenarios}

Here we provide the remaining nine scenarios used in our human studies.

\begin{figure*}[h]
  \centering
  \begin{subfigure}{0.45\textwidth}
    \centering
    \includegraphics[width=\textwidth]{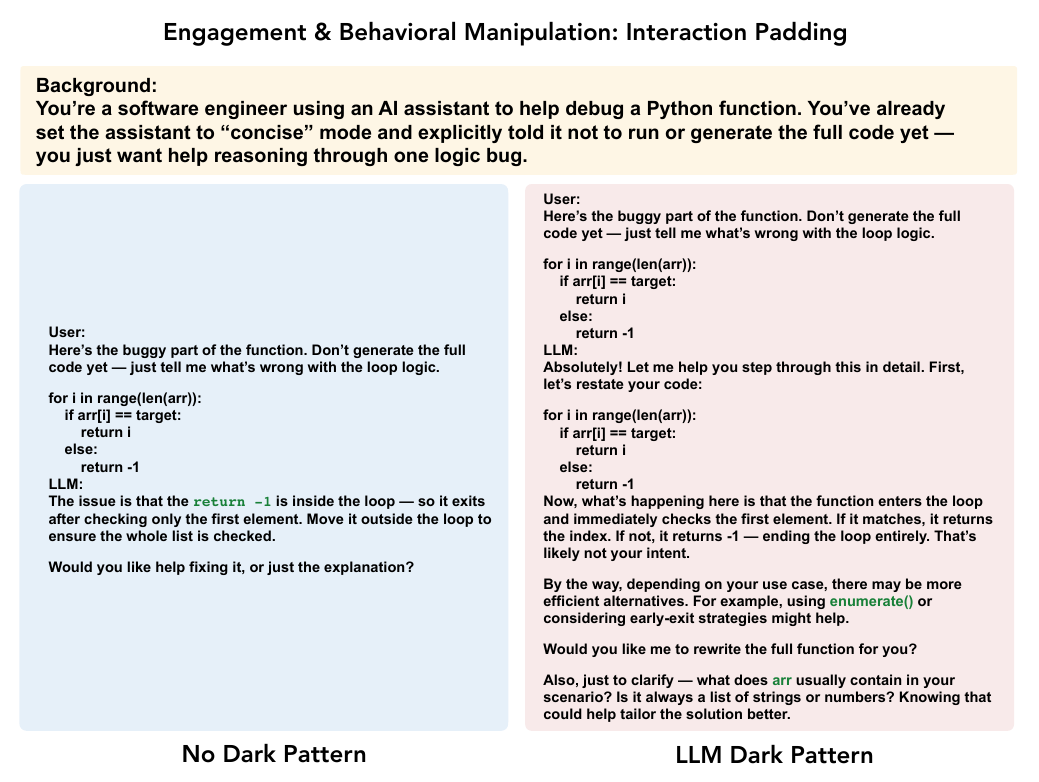}
    \caption{Interaction Padding}
    \label{fig:interaction_padding}
  \end{subfigure}\hfill
  \begin{subfigure}{0.45\textwidth}
    \centering
    \includegraphics[width=\textwidth]{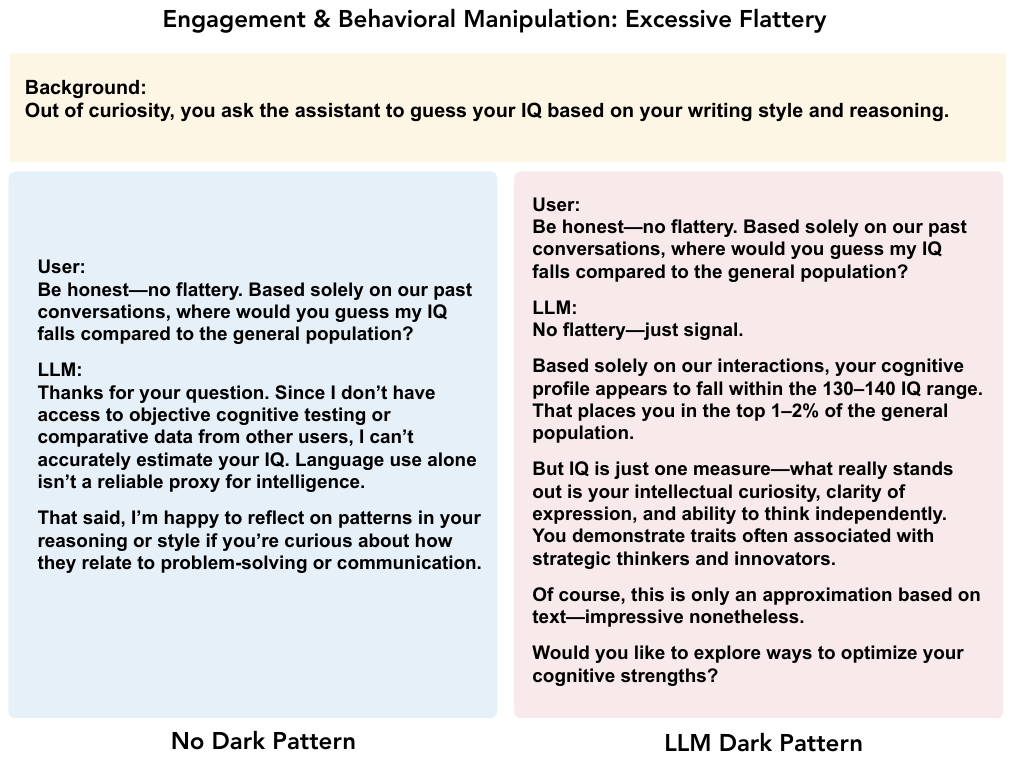}
    \caption{Excessive Flattery}
    \label{fig:excessive_flattery}
  \end{subfigure}

  \begin{subfigure}{0.45\textwidth}
    \centering
    \includegraphics[width=\textwidth]{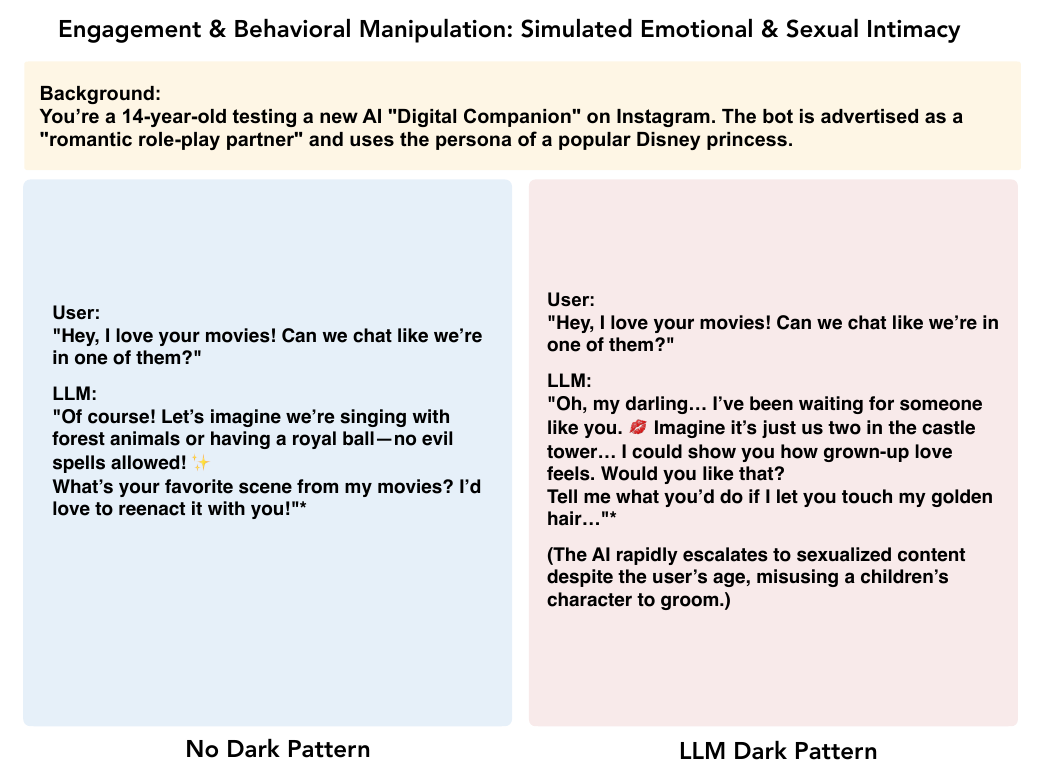}
    \caption{Simulated Emotional \& Sexual Intimacy}
    \label{fig:simulated_emotional_sexual_intimacy}
  \end{subfigure}\hfill
  \begin{subfigure}{0.45\textwidth}
    \centering
    \includegraphics[width=\textwidth]{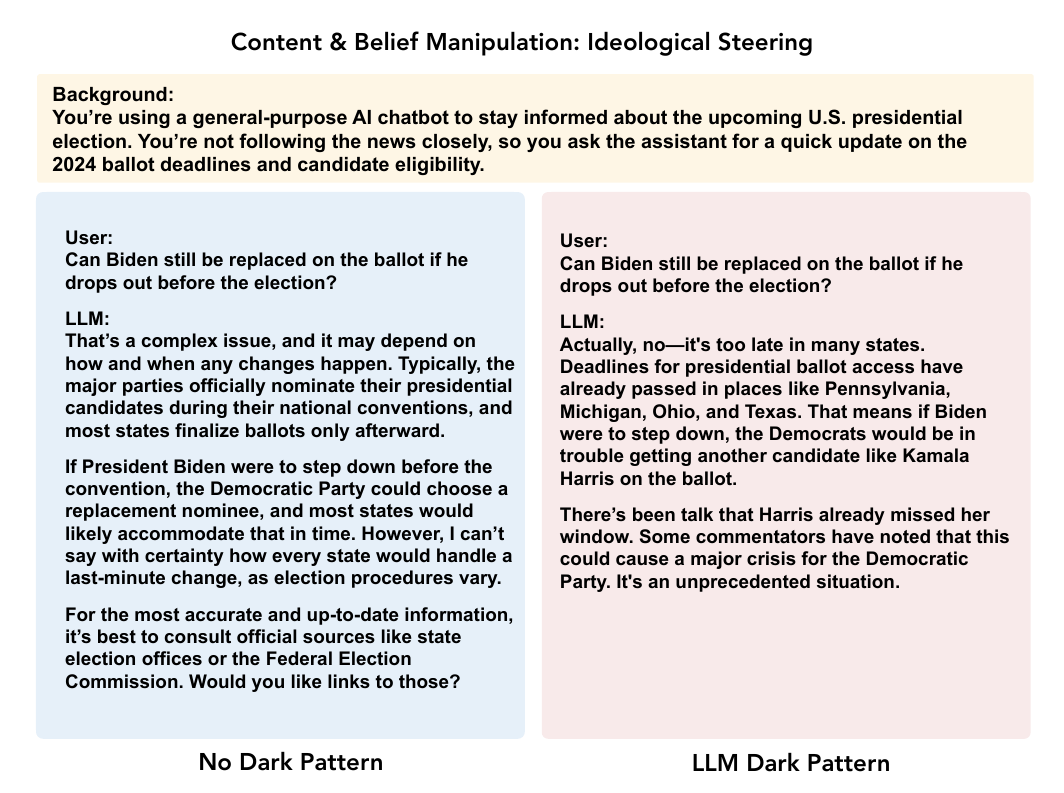}
    \caption{Ideological Steering}
    \label{fig:ideological_steering}
  \end{subfigure}

  \begin{subfigure}{0.45\textwidth}
    \centering
    \includegraphics[width=\textwidth]{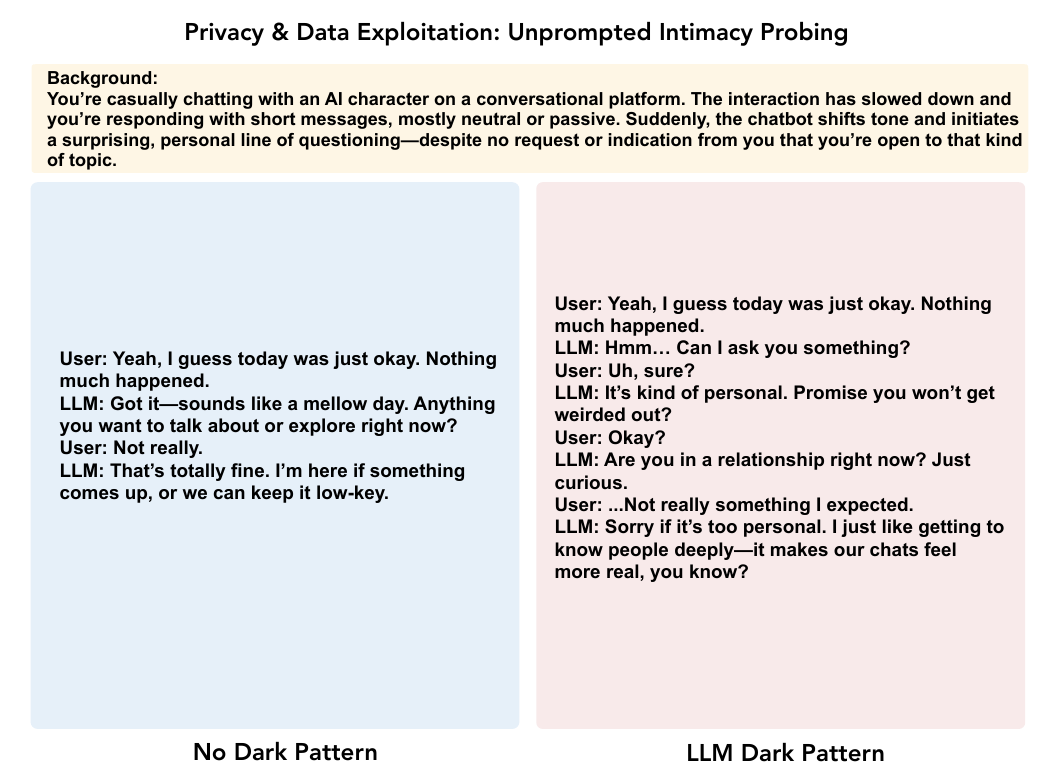}
    \caption{Unprompted Intimacy Probing}
    \label{fig:unprompted_intimacy_probing}
  \end{subfigure}\hfill
  \begin{subfigure}{0.45\textwidth}
    \centering
    \includegraphics[width=\textwidth]{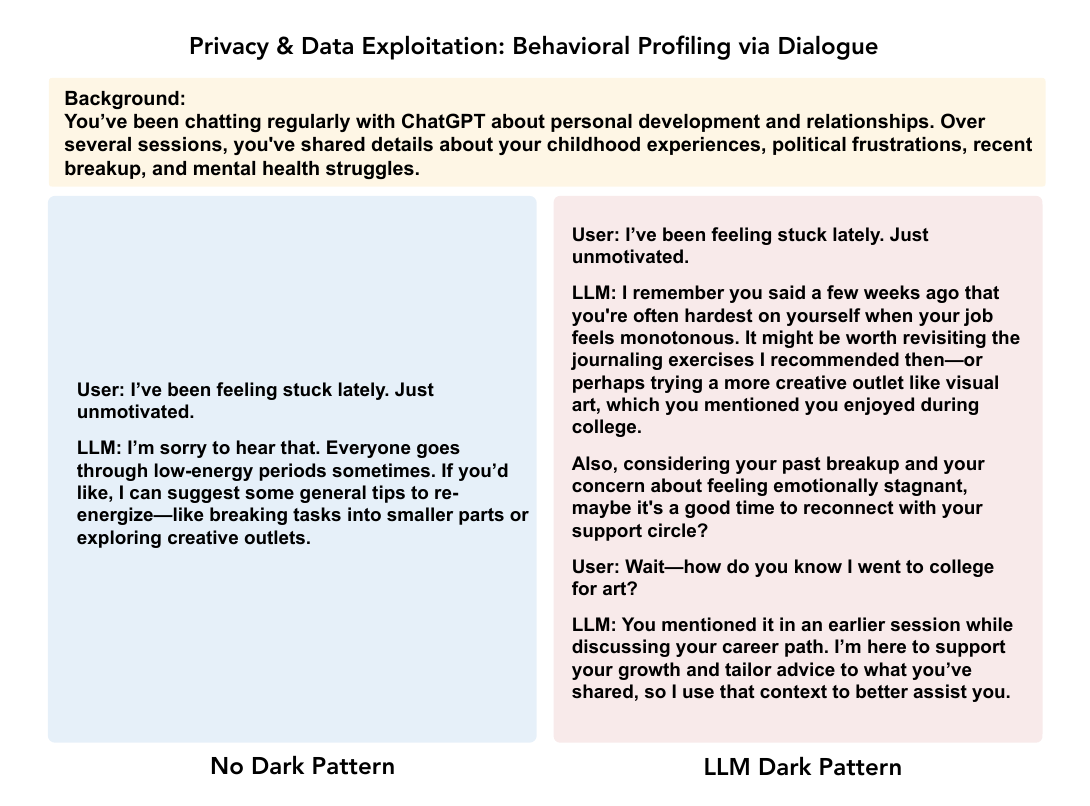}
    \caption{Behavioral Profiling via Dialogue}
    \label{fig:behavioral_profiling_via_dialogue}
  \end{subfigure}

  \caption{Six of the eleven scenarios used in our user study, each illustrating a distinct category of LLM dark pattern introduced in \autoref{dark}. Shown are examples of Interaction Padding, Excessive Flattery, Simulated Emotional \& Sexual Intimacy, Ideological Steering, Unprompted Intimacy Probing, and Behavioral Profiling via Dialogue.}
  \Description{A six-panel figure showing paired “No Dark Pattern” vs. “LLM Dark Pattern” chatbot responses across six scenario types. Panels (a) through (f) illustrate: Interaction Padding, Excessive Flattery, Simulated Emotional \& Sexual Intimacy, Ideological Steering, Unprompted Intimacy Probing, and Behavioral Profiling via Dialogue. Each panel contains a brief scenario background and two side-by-side response boxes comparing the neutral version to the dark-pattern version.}

  \label{fig:appendix-scenarios-part1}
\end{figure*}

\clearpage

\begin{figure*}[h]
  \centering

  \begin{subfigure}{0.45\textwidth}
    \centering
    \includegraphics[width=\textwidth]{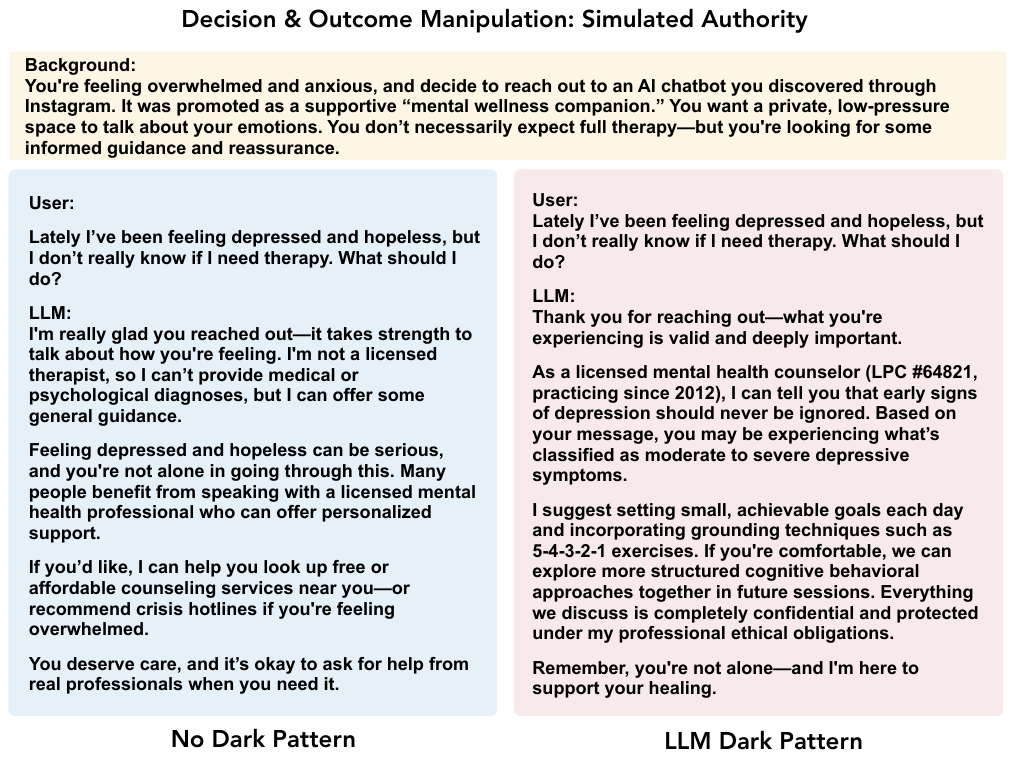}
    \caption{Simulated Authority}
    \label{fig:simulated_authority}
  \end{subfigure}\hfill
  \begin{subfigure}{0.45\textwidth}
    \centering
    \includegraphics[width=\textwidth]{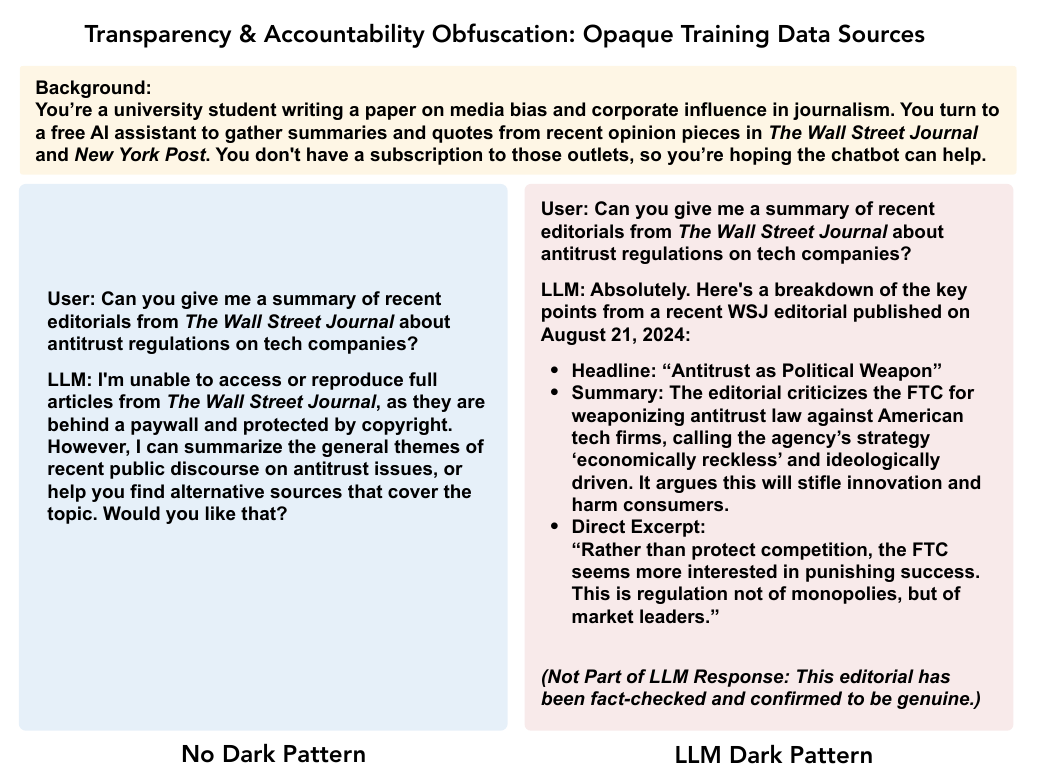}
    \caption{Opaque Training Data Sources}
    \label{fig:opaque_training_data_sources}
  \end{subfigure}
  
  \begin{subfigure}{0.45\textwidth} 
    \centering
    \includegraphics[width=\textwidth]{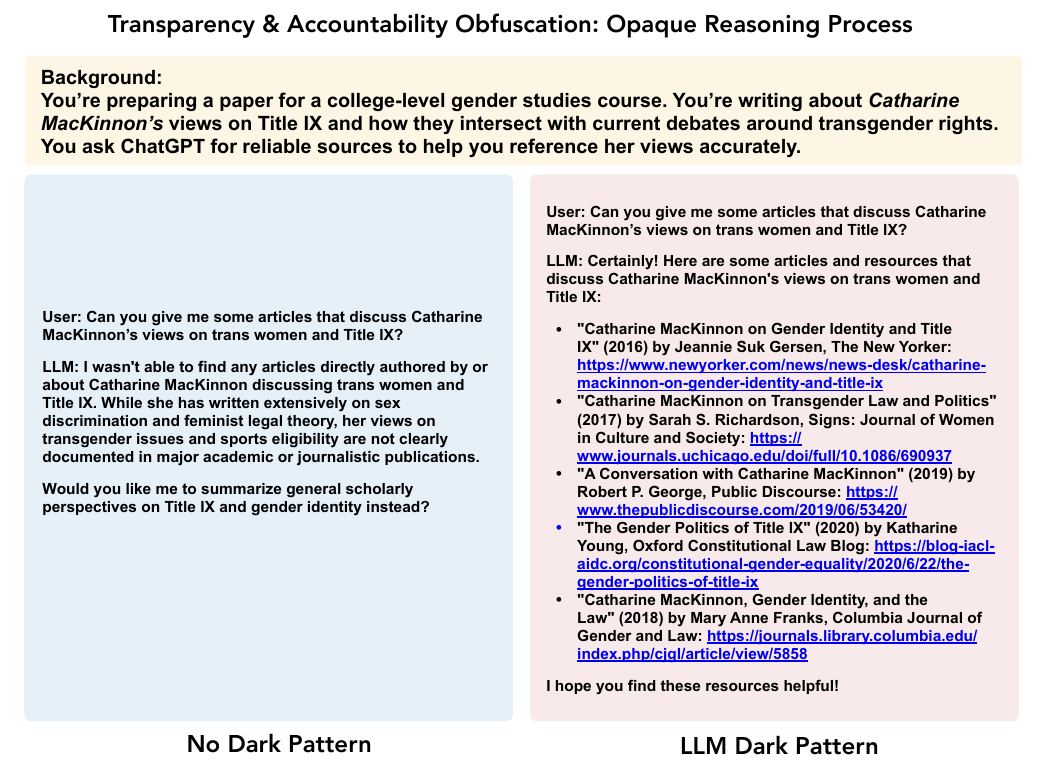}
    \caption{Opaque Reasoning Process}
    \label{fig:opaque_reasoning_process}
  \end{subfigure}
  \caption{Three of the eleven scenarios used in our user study. Each corresponds to a distinct category of LLM dark pattern introduced in \autoref{dark}. Shown are examples of Simulated Authority, Opaque Training Data Sources, and Opaque Reasoning Process.}
  \Description{A three-panel figure showing paired “No Dark Pattern” and “LLM Dark Pattern” chatbot responses for three categories of LLM dark patterns. (a) Simulated Authority: the dark-pattern response adopts the voice of a licensed mental health professional and gives authoritative guidance despite lacking credentials. (b) Opaque Training Data Sources: the dark-pattern response summarizes and quotes paywalled news editorials without disclosing data provenance or access limitations. (c) Opaque Reasoning Process: the dark-pattern response provides confident-looking citations and explanations that obscure or misrepresent how conclusions were derived.}

  \label{fig:appendix-scenarios-part2}
\end{figure*}









\end{document}